\documentclass[acmsmall]{acmart}
\usepackage{array}
\usepackage{multirow}
\usepackage{xcolor}
\usepackage{svg}
\usepackage{pdfpages}
\usepackage{threeparttable}
\usepackage{colortbl}
\usepackage{hyperref}
\usepackage{svg}
\usepackage{fontawesome}
\usepackage{tcolorbox}
\usepackage{mdframed}
\usepackage{enumitem}

\definecolor{mygray}{gray}{0.25}
\definecolor{mygray2}{gray}{0.8}
\definecolor{mygray3}{gray}{0.9}
\definecolor{amber}{rgb}{1.0, 0.75, 0.0}
\definecolor{mygreen}{rgb}{0.0, 0.5, 0.0}

\AtBeginDocument{%
  \providecommand\BibTeX{{%
    \normalfont B\kern-0.5em{\scshape i\kern-0.25em b}\kern-0.8em\TeX}}}

\setcopyright{acmlicensed}
\acmJournal{PACMHCI}
\acmYear{2021} \acmVolume{5} \acmNumber{CSCW2} \acmArticle{348} \acmMonth{10} \acmPrice{}\acmDOI{10.1145/3476089}

\received{January 2021}
\received[revised]{April 2021}
\received[accepted]{May 2021}

\newenvironment{myquote}%
  {\list{}{\leftmargin=0.2in\rightmargin=0.2in}\item[]}%
  {\endlist}

\begin{document}

\title{A Framework of High-Stakes Algorithmic Decision-Making for the Public Sector Developed through a Case Study of Child-Welfare}

\author{Devansh Saxena}
\affiliation{%
  \institution{Marquette University}
  \streetaddress{Cudahy Hall, 1313 W Wisconsin Avenue}
  \city{Milwaukee}
  \state{WI}
  \postcode{53233}
  \country{USA}}
\email{devansh.saxena@marquette.edu}

\author{Karla Badillo-Urquiola}
\affiliation{%
  \institution{University of Central Florida}
  \streetaddress{4000 Central Florida Blvd}
  \city{Orlando}
  \state{FL}
  \postcode{32816}
  \country{USA}}
\email{kcurquiola10@knights.ucf.edu}

\author{Pamela J. Wisniewski}
\affiliation{%
  \institution{University of Central Florida}
  \streetaddress{4000 Central Florida Blvd}
  \city{Orlando}
  \state{FL}
  \postcode{32816}
  \country{USA}}
\email{pamwis@ucf.edu}

\author{Shion Guha}
\affiliation{%
  \institution{University of Toronto}
  \streetaddress{140 St. George Street}
  \city{Toronto}
  \state{Ontario}
  \country{Canada}}
\email{shion.guha@utoronto.ca}

\renewcommand{\shortauthors}{Devansh Saxena et al.}

\begin{abstract}
Algorithms have permeated throughout civil government and society, where they are being used to make high-stakes decisions about human lives. In this paper, we first develop a cohesive framework of algorithmic decision-making adapted for the public sector (ADMAPS) that reflects the complex socio-technical interactions between \textit{human discretion}, \textit{bureaucratic processes}, and \textit{algorithmic decision-making} by synthesizing disparate bodies of work in the fields of Human-Computer Interaction (HCI), Science and Technology Studies (STS), and Public Administration (PA). We then applied the ADMAPS framework to conduct a qualitative analysis of an in-depth, eight-month ethnographic case study of algorithms in daily use within a child-welfare agency that serves approximately 900 families and 1300 children in the mid-western United States. Overall, we found that there is a need to focus on strength-based algorithmic outcomes centered in social ecological frameworks. In addition, algorithmic systems need to support existing bureaucratic processes and augment human discretion, rather than replace it.  Finally, collective buy-in in algorithmic systems requires trust in the target outcomes at both the practitioner and bureaucratic levels. As a result of our study, we propose guidelines for the design of high-stakes algorithmic decision-making tools in the child-welfare system, and more generally, in the public sector. We empirically validate the theoretically derived ADMAPS framework to demonstrate how it can be useful for systematically making pragmatic decisions about the design of algorithms for the public sector. 

\end{abstract}



\maketitle

\section{Introduction}
The influence of neoliberal politics and theories of New Public Management (NPM) \cite{lane2000new} throughout most modern societies over the past two decades has sought to reform public services by emulating corporations to improve efficiency \cite{legreid2017transcending}. One way to achieve this goal for public sector services (e.g., child-welfare, labor, criminal justice, and public education) is through the adoption of automated processes (e.g., decision-making algorithms), as they purportedly promise to increase efficiencies, lower costs, and provide better outcomes for citizens. \cite{ringel2018improving}. As such, algorithms in the public sector have become pervasive and, in turn, well-studied in recent years \cite{maciejewski2017more, eubanks2018automating, christin2015courts}. Consequently, they have also been scrutinized for achieving worse outcomes, exacerbating racial biases, and strengthening structural inequalities \cite{eubanks2018automating, redden2020datafied, saxena2020human} within systems that are overburdened and under-resourced, yet critically needed \cite{lobao2018shrinking, davidson2020extreme}.   

As a case in point, 423,997 children were in the U.S. Child Welfare System (CWS) in September 2019 which represents a steady increase in the past decade \cite{us2019afcars}. This number is only expected to grow in upcoming years barring major structural reforms. This has created an ever increasing burden for CWS workers to make decisions about children that provide positive outcomes for them. Policymakers have decided that one avenue to address this issue is to implement algorithmic decision-making within CWS \cite{congressbill}. As such, algorithmic decision-making tools are now being used in high-stakes CWS situations, including making risk assessments of child abuse \cite{de2020case} and determining placement stability \cite{moore2016assessing}. Brown et al. \cite{brown2019toward} conducted co-design workshops with stakeholders within the CW community (e.g., families, frontline providers, and specialists) and found that such algorithms largely bolstered distrust, perpetuated bias, and created black-boxed systems, which accelerated concerns about how these tools may negatively impact child-welfare workers’ decisions. Thus, any technological solution cannot be inherently deemed `fair' or `just' and require complementary policy changes to affect community perceptions \cite{brown2019toward}. In a recent comprehensive review of the literature, Saxena et al. \cite{saxena2020human} highlighted the lack of human-centeredness \cite{baumer2017toward} in the design and implementation of these algorithms and the need for more empirical work on how these algorithms are embedded in the daily work practices of child-welfare caseworkers. Thus, the gaps identified in these prior works led to the following over-arching research questions:
\begin {itemize} [leftmargin=*]
  \item \textbf{RQ1:} \textit{What are the high-stakes outcomes for which algorithmic decision-making is leveraged within the child-welfare system?}
  \item \textbf{RQ2:} \textit{How does the implementation of a given algorithm impact algorithmic decision-making, human discretion, and bureaucratic processes?}
  \item \textbf{RQ3:} \textit{What are the potential benefits and drawbacks when balancing the trade-offs between these three elements?}
\end{itemize}

To address these questions, first, we synthesized prior literature to develop a theoretical framework for Algorithmic Decision-Making Adapted for the Public Sector (ADMAPS). While SIGCHI researchers have attempted to formalize the dimensions of algorithmic decision-making in various contexts \cite{alkhatib2019street, paakkonen2020bureaucracy, holten2020shifting}, we argue that the high-stakes decisions being made within the public sector necessitate the critical need for a distinctly unique framework for algorithmic systems that accounts for the complexities of public sector bureaucratic processes \cite{farazmand2009bureaucracy, loyens2010toward, frederickson2015public} and the delicate application of human discretion that has historically been a cornerstone of social services \cite{lipsky2010street, sosin2010discretion}. We did this by synthesizing related works from the fields of Human-Computer Interaction (HCI), Science and Technology Studies (STS), and Public Administration (PA).

Next, we leveraged the ADMAPS framework as a theoretical lens in which to analyze the qualitative data collected from an eight-month in-depth ethnographic case study of a child-welfare agency in the mid-western United States. We attended 55 agency meetings and conducted 20 individual interviews over the course of eight months, which resulted in daily interactions with approximately 120 CWS agency employees and external consultants. To answer (\textbf{RQ1}), we first identified the algorithms used within the CW agency and the relevant data and outcomes they addressed within the system. For (\textbf{RQ2}), we assessed whether and how each algorithm affected each dimension of \textbf{human discretion} (\textit{professional expertise, value judgments, heuristic decision-making}), \textbf{bureaucratic processes} (\textit{resources and constraints, administration and training, laws and policies}), and \textbf{algorithmic decision-making} (\textit{relevant data, types of decision-support, degree of uncertainty}). Finally, for (\textbf{RQ3}), we compared the four algorithms identified (i.e., CANS, 7ei, AST, and LPS) to show how ADMAPS can help balance the trade-offs in algorithmic decision-making to optimize the benefits and minimize the drawbacks associated with the high-stakes outcomes within the CWS.   

Overall, we found that there is a need to refocus on strength-based outcomes centered in social ecological frameworks \cite{bronfenbrenner1975reality} \textbf{(RQ1)}. We define strength-based outcomes as those that draw upon a person's assets and strengths rather than their deficits and weaknesses \cite{badillo2018chibest, zimmerman2013resiliency}. In addition, an over-reliance on algorithmic decision-making to support bureaucratic processes can be detrimental to human discretion, however, algorithmic decision-making can support human discretionary work if they are fully supported by bureaucratic processes \textbf{(RQ2)}. Finally, algorithmic decision-making should augment human discretion (by building theory-driven algorithms centered in practice) rather than attempt to replace it; algorithm decision-support systems and bureaucratic processes need to be aligned (lack of balance creates utility issues) and collective buy-in in such systems requires trust in algorithmic outcomes at both the caseworker and bureaucratic levels \textbf{(RQ3)}. Thus, this paper makes the following unique research contributions:


\begin {itemize}[leftmargin=*]
  \item We conducted an in-depth ethnographic case study to understand daily algorithmic decision-making practices of caseworkers in CWS. 
  \item We go beyond existing recommendations for AI/ML to provide specific heuristic guidelines for algorithmic decision-making in CWS that can be of use to caseworkers, supervisors, government bureaucrats and policymakers.
  \item We developed a theoretical framework (ADMAPS) of algorithmic decision-making in the public sector that synthesizes prior work on algorithmic decision-making in non-public sector settings with the unique challenges, limitations and opportunities in the public sector. ADMAPS is generalizable to a wider range of public sector domains such as the criminal justice system,  unemployment services, and public education. 
\end{itemize}

In the sections that follow, we first highlight some of the the high-stakes decisions made within the public sector. Then, we introduce our framework for \textbf{Algorithmic Decision-Making Adapted for the Public Sector (ADMAPS)}. Finally, we use ADMAPS to present an in-depth ethnographic case study of four algorithms used daily in CWS to determine high-stakes outcomes for foster children, including trauma-informed care, placement stability, and sex-trafficking risk.  

\section{The High-Stakes Decisions Made within the Public Sector}
Algorithmic systems are being used to make high-stakes decisions about human lives and welfare in the public sector ranging across child-welfare, criminal justice, public education, job placement centers, welfare benefits, and housing among others. For example, the criminal justice system employs algorithms to determine length of sentencing \cite{grgic2019human}, allocate resources to neighborhoods \cite{chainey2013examining}, and predict the likelihood of recidivism (i.e., recommitting a crime) \cite{drawve2019utilizing}. In the child-welfare system, decisions are being made about whether to remove a child from the care of their parents based on the risk of future maltreatment \cite{cuccaro2017risk}, who should be raising a child \cite{moore2016assessing}, and what types of services should be offered to families \cite{gillingham2018decision}. The public education system also uses algorithms to assign students to public school zones \cite{robertson2020if, robertson2021modeling} and determine student performance \cite{williamson2016digital}. Job placement centers profile job seekers and make job placement decisions using algorithms as well \cite{allhutter2020algorithmic, holten2020shifting}. Algorithms are also used to establish eligibility criteria for receiving benefits and offer these benefits to families in need \cite{eubanks2018automating}. In short, many of the ways in which algorithms are being implemented in the public sector result in life altering, if not life and/or death consequences.  

The public sector differs uniquely from the private sector in terms of algorithmic decision-making in two distinct ways. First, outcomes in the public sector like assessing risk of recidivism in criminal justice or assessing need in welfare services are poorly and inconsistently defined \cite{greene2020hidden, barocas2016big, saxena2020human, eubanks2018automating, zgoba2015recidivism}. Moreover, an individual's personal situation can (de)stabilize several times making it hard to assess what constitutes a successful outcome or intervention \cite{holten2020shifting}. Second, current practices of using aggregate administrative data that is often poorly collected \cite{connelly2016role, saxena2020human, gillingham2018decision} and biased to predict an individual's behavior is a complex and hard task that may lead to unfairness in decision-making outcomes and is, in most western, liberal democratic systems unconstitutional and/or illegal \cite{voigt2017eu}. These two factors combine to make algorithmic decision-making in the public sector, a high stakes decision-making environment that has real repercussions for the lives and liberties of people. Therefore, the algorithmic decision-making process in the public sector needs to be scrutinized with the utmost care. Thus, there is an urgent need to develop a cohesive, yet tailored framework for algorithmic decision-making within the public sector that is validated with in-depth, empirical work focusing on daily algorithmic practices and decision-making. 

\vspace{-0.1cm}
\subsection{Algorithmic Decision-Making within the Child-Welfare System}
Algorithms in child-welfare have historically relied on a narrow set of psychometric predictors that are used to assess the risks and needs of foster children and parents. However, a more comprehensive understanding about the accumulation of risk is necessary to account for the family’s social support system as well as the risk posed by the system itself \cite{saxena2020human}. Saxena et al. \cite{saxena2020human} recently conducted a systematic review of algorithms employed in the U.S. child-welfare system and uncovered several discrepancies in regard to data, computational methods, and target outcomes of these algorithms \cite{saxena2020human}. There is a need for theoretically constructed algorithms centered in the nature of practice. Moreover, decisions must be made within the constraints of policies and systemic barriers; characteristics not accounted for by algorithms. The majority of algorithms in CWS are empirically constructed even though the empirical knowledge in child-welfare is quite fragmented and social science theories are needed to fill in these gaps \cite{gambrill2000risk}. For instance, child-welfare workers are often frustrated by algorithms because they do not account for the scarce resources in the public sector \cite{gambrill2001need}. Risk assessment has also been the dominant focus of algorithms in CWS, however, there are concerns about their deficit-based nature, which only seeks to minimize risk but not improve the quality of children’s lives. This is driving attention away from strength-based frameworks \cite{saxena2020human, brown2019toward, pinter2017adolescent, badillo2018chibest}. Prior work has also explored the utility of algorithms designed to aid decision-making and found that they increased uncertainty and led to unreliable decisions since caseworkers were required to translate information from both the clinical and algorithmic assessments \cite{shlonsky2005next, schwalbe2004re}. More recently, researchers have also focused on the need to uncover politics, economics, and social implications of CWS algorithms and established the need to actively work with practitioners and domain experts to understand their perspectives about such systems as well as the systemic factors centered in policies, laws, and organizational culture that play a significant role in decision-making \cite{saxena2020conducting, redden2020datafied, gillingham2019decision, brown2019toward}. This paper responds to these calls by a conducting a deep ethnographic analysis of algorithms in use at a CWS agency and uncovers their social, technical, and political implications.

\vspace{-0.1cm}
\section{A Framework of Algorithmic Decision-Making Adapted for the Public Sector (ADMAPS)}
As shown in Figure 1, we leveraged a socio-technical perspective of algorithmic decision-making that captures the three-way interactions between: 1) \textbf{human discretion}, 2) \textbf{bureaucratic processes}, and 3) \textbf{algorithmic decision-making}. We did this by synthesizing relevant, yet disparate, bodies of work across the fields of Public Administration (PA), Science and Technologies Studies (STS), and Human-Computer Interaction (HCI) to create a cohesive framework of Algorithmic Decision-Making Adapted for the Public Sector (ADMAPS). This framework is a core contribution of this paper and also served as a theoretical lens for grounding the qualitative analyses of our empirical case study within the domain of child-welfare. 

\begin{figure}
  \includegraphics[scale=0.25]{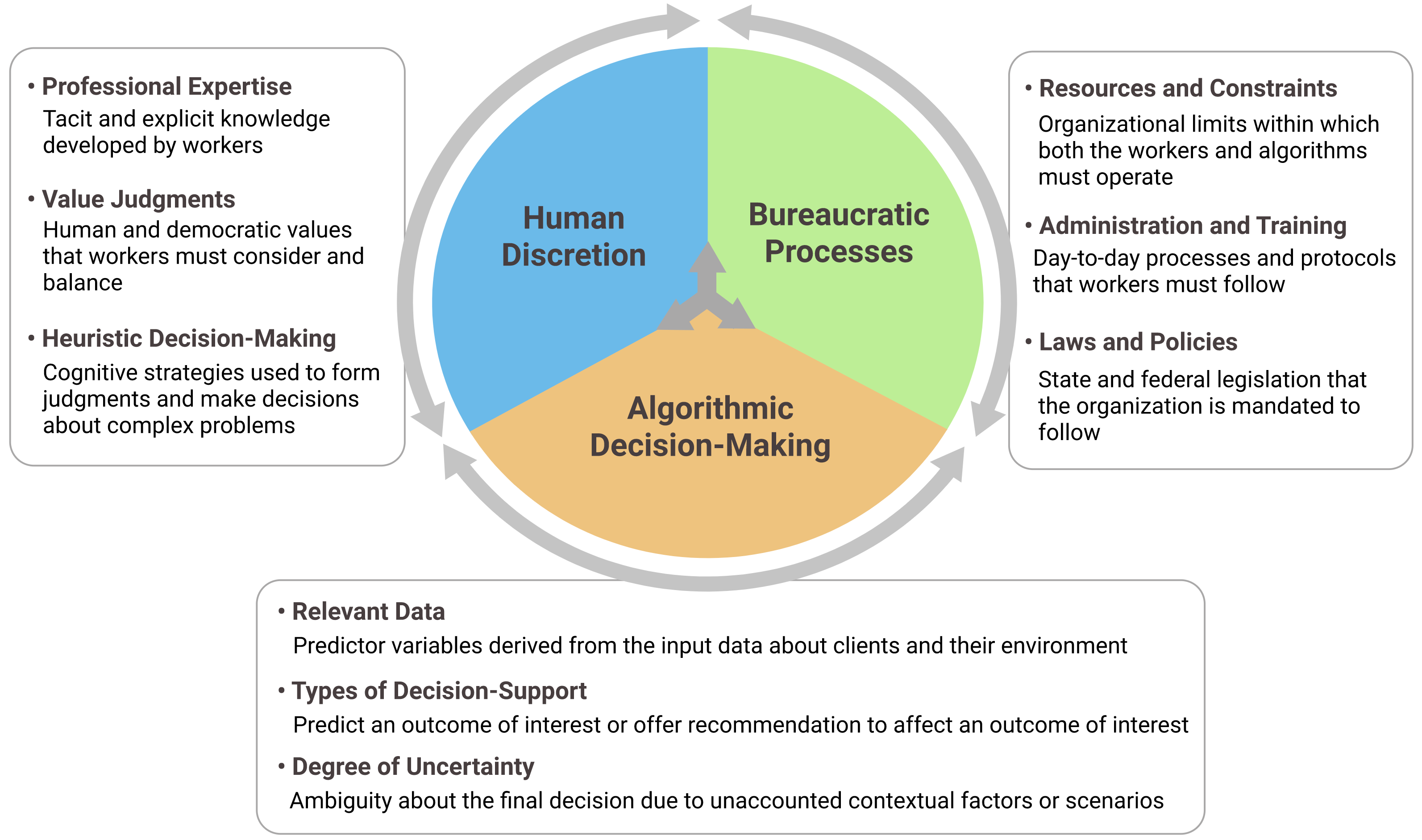}
  \caption{A Framework for Algorithmic Decision-Making Adapted for the Public Sector (ADMAPS)}
  \vspace{-0.5cm}
  \label{fig:framework}
\end{figure}

Scholars within the PA field have extensively studied how \textbf{human discretion} by street-level bureaucrats \footnote{A street-level bureaucrat is a professional service worker (e.g., social worker, police officer, teacher) who operates in the frontline of public service provision. They interact closely with clients and makes decisions about them based on how they interpret policies relating to the situations at hand \cite{lipsky2010street}.} plays a central role in navigating \textbf{bureaucratic processes} and implementing policies. For example, bureaucrats act with a certain level of autonomy in how they interpret and apply professional standards when determining which clients must receive welfare benefits or services \cite{sosin2010discretion}. Public Administration scholars have also started to recognize the impact of information communication technologies (ICTs) on the nature of human discretion and bureaucracy with some recent attention paid to artificial intelligence \cite{busch2018digital, peeters2020agency, m2020artificial, bullock2019artificial, young2019artificial}. Young et al. \cite{young2019artificial} introduced \textit{artificial discretion} as a theoretical framework to help public managers assess the impact of AI and how it differs from human discretion with respect to task specificity and environmental complexity. However, several of these studies, rich in their understanding of human discretion and bureaucracy, continue to treat algorithms as peripheral end products; a new part of bureaucracy to which human discretion must adapt. Meanwhile, STS scholars conducting studies in the public sector have used Kitchin and Lauriault's framework of \textit{data assemblages} \cite{kitchin2014towards} to deeply study the intersection of \textbf{bureaucratic processes} and \textbf{algorithmic decision-making} by examining the politics of data systems \cite{redden2020datafied, williamson2016digital, leszczynski2016speculative}. Data assemblages perceive data systems as complex assemblages of human actors, artifacts, technical systems, institutions and ideas. This framing provides a means to consider how these systems are socially, economically, and politically constructed. Similarly, STS scholars have also used Seaver’s notion of \textit{algorithms as culture} \cite{seaver2017algorithms} to understand the social implications and values of algorithmic systems through an ethnographic analysis. These studies have made significant contributions towards the community’s understanding of how algorithms shape and are shaped by cultural context, how value is inscribed to algorithms, and how power is afforded to them. However, the societal and cultural perspective applied within field of STS may at times miss some of the nuance at the bureaucratic street-level, which is where human discretion is most critical. Due to the high-stakes decisions being made within the public sector, there is an urgent need to map the complex interdependencies between the core elements of human discretion, bureaucratic processes, and algorithmic decision-making that often go unnoticed in the public sector. 

The SIGCHI research community is well-positioned to do this cross-disciplinary and integrative work due to our strengths in taking a human-centered \cite{baumer2017toward} and value-sensitive approach \cite{zhu2018value} to the design and development of algorithms. Furthermore, CSCW is a well-suited venue for this type of research due to the collaborative nature of the work being performed by teams of CWS employees and external consultants when making critical decisions about the well-being of children. In the sections below, we describe the key dimensions of our framework.

\vspace{-0.1cm}
\subsection{Human Discretion}
Lipsky's theory of street-level bureaucrats \cite{lipsky2010street} defines \textit{human discretion} as an individual’s ability to exercise their own judgment in implementing government policies in complex and uncertain problem spaces. Scholars across multiple disciplines have recognized the importance of human discretion in developing algorithms and interpreting and administering algorithmic decisions \cite{baumer2017toward, passi2017data, passi2018trust, passi2020making, dourish2018datafication} but also in interpreting and making policy decisions \cite{lipsky2010street, chalekian2013posdcorb, frederickson2015public, bullock2018sector}. However, when HCI scholars deliberate over human discretion, it generally occurs from a design perspective. That is, how can we incorporate humans' tacit knowledge, social interpretations, and values into the design process \cite{baumer2017toward, zhu2018value}. Whereas, when Public Administration scholars discuss human discretion, it refers to the decision-making latitude as well as the value-laden choices that bureaucrats must make when experiencing complexity and uncertainty \cite{lipsky2010street, busch2018digital, bullock2018sector}. We integrate this knowledge about human discretion and close this loop by presenting the following process model where the bureaucrats use their professional expertise, engage in value judgments, and heuristic decision-making. 

\vspace{-0.1cm}
\subsubsection{Professional Expertise} The tacit and explicit knowledge developed by workers in any given domain \cite{baumer2017toward, smith2001role}. It plays an important role with respect to the workers' confidence in their own decisions \cite{pillay2005distributed} as well as the level of adeptness with which they navigate bureaucratic processes \cite{frederickson2015public, chalekian2013posdcorb}, negotiate resources \cite{bullock2018sector}, and seek additional supervision \cite{ericsson1991toward}. As novice practitioners gain confidence in their skills, they become more aware of the need to pursue additional details, supervision, and other opportunities to increase their level of domain knowledge \cite{welch2008metasynthesis}. Professional expertise, however, is also domain-specific and continually evolves with time \cite{pillay2005distributed, ericsson1991toward}. It is necessary to examine the nature professional expertise within the public sector which is rapidly evolving through the continued digitization and automation of work processes that were previously the forte of street-level bureaucrats \cite{busch2018digital, giest2018unraveling}. Practitioners in the public sector are continually acquiring new skills as they learn to make decisions through data systems and interpret algorithmic outputs \cite{bovens2002street, bullock2019artificial}, however, these new skills are not being accounted for with respect to what constitutes professional expertise \cite{holten2020shifting}.

\subsubsection{Value Judgments}
Practitioners must consider and balance human and democratic values when assessing cases about citizens \cite{holten2020shifting, friedman2002value}. This is another key dimension of human discretion because workers must weigh the competing motivations of different clients as well as differing notions of values and incorporate them within their decision-making processes. Value judgments play a pertinent role when a practitioner is faced with ethical dilemmas and informs their decision-making ability. Practitioner's beliefs and moral values are important factors in regard to how street-level decision-making unfolds \cite{maynard2003cops, sandfort2000moving, loyens2010toward}. Moreover, practitioner's personal values are often mediated by organizational culture which subsequently yields results that can be significantly different than results centered in personal values \cite{fritzsche1991model}. Therefore, it is essential to understand the role of value judgments vis-a-vis human discretion because practitioners have assumed the role of value mediators who must weigh the needs of citizens against the demands of policymakers \cite{pors2020street}. 

\subsubsection{Heuristic Decision-Making}
Heuristics refer to the cognitive strategies used to form judgments, make decisions, and find solutions to complex problems \cite{gigerenzer2011heuristic}. Gigerenzer and Gaissmaier \cite{gigerenzer2011heuristic} reviewed research on heuristic decision-making in business organizations, health care, and legal institutions and established the fundamental role it plays within organizations. Heuristic decision-making can lead to more accurate decisions than complex rational models and selecting information in an adaptive manner can lead to more accurate judgments than weighing all of the information \cite{mousavi2014risk}. Practitioners work more effectively and efficiently when their knowledge base in well-organized and centered in heuristics since it allows them to separate relevant and irrelevant knowledge for any given context \cite{arts2000expertise}. Simple heuristics can be more successful especially in uncertain and complex spaces since all the information required to make a decision might not be available as a result of uncertainty \cite{artinger2015heuristics}. Therefore, it becomes imperative for practitioners to rely on their heuristics that are acquired through experience and practice \cite{welch2008metasynthesis}. Moreover, decision-making in organizations must involve professionals' heuristics because the ideal conditions required for rational and reductive models rarely hold true in an uncertain world \cite{payne1993use}. 

\subsection{Bureaucratic Processes} 
Bureaucratic processes are the critical governance characteristics essential for policy development and implementation to serve public interests \cite{farazmand2009bureaucracy, farazmand2017governance, frederickson2015public}. Both HCI and Public Administration scholars have recognized the dominant role that bureaucratic processes play both with respect to establishing the role of bureaucrats (i.e., human discretion) as well as the adoption of technology \cite{bovens2002street, bullock2018sector, lodato2018institutional, saxena2020conducting}. Scholars have also emphasized that policy/bureaucratic considerations must precede technology design and professional practice considerations \cite{jackson2014policy}. We include three different dimensions to bureaucratic processes consistently highlighted in the literature as described below.  

\subsubsection{Resources and Constraints} 
Availability of essential resources (administrative, financial, personnel, political) directly impact organizational performance \cite{lee2013assessing, fredericksen2000disconnect, manzoor2014look}. Resources can also be viewed as constraints within which the organization must operate \cite{lee2013assessing}. This dimension is of special importance for the public sector that is facing severely limited resources and new dilemmas in the form of burdensome workloads, high staff turnover, and a lack of experienced workers \cite{manzoor2014look, milakovich2013public}. Examining how these scare resources are allocated in public services is crucial because most agencies are experiencing a push to innovate and invest in evidence-based practices to improve performance \cite{vigoda2011change}, however, investing in innovation can be challenging in a resource-deficit domain \cite{osborne2013handbook}.

\subsubsection{Administration and Training} 
Protocols, workflows, and processes established at the organization that are followed by workers in their day-to-day practice and play a significant role in decision-making \cite{lipsky2010street, lee2013assessing, peters2014accountability}. Organizational processes offer the means to understand how an agency makes decisions within policy mandates as well as how it meets diverse public needs \cite{boschken1994organizational}. Processes are established to improve accountability in the form of consistent, transparent, and defensible decision-making and allow the agency to effectively communicate compliance with legal mandates as well as utilize existing knowledge routines to improve reliability \cite{peters2014accountability, dean1993procedural}. Moreover, processes followed by practitioners in their daily lives also continually shape policy on the ground \cite{lipsky2010street, brodkin2008accountability}. This dimension also identifies the workers' training in the public sector which plays a critical role in regard to individual, team, and organizational development \cite{kennett2013impact}. New assessments and tools are continually being introduced in the public sector in a pursuit for creativity and innovation such that it leads to standardized and evidence based decision-making \cite{desmarchelier2019innovation, bullock2019artificial, carter2018building}, however, this also necessitates a need to examine if workers are being adequately trained to fully utilize these tools \cite{van2021training}. Training at the organization also establishes the basis for worker expertise by ensuring that the workers skillfully mediate both the nature of practice and bureaucratic processes \cite{welch2008metasynthesis}. 

\vspace{-0.1cm}
\subsubsection{Law and Policies} Formal actions enacted by legislatures or political executives that public administrators must comply with and implement \cite{peeters2020agency}. Laws and policies establish the constraints within which all decisions (human or algorithmic) must be made as well as define the outcomes of interests themselves. For instance, law dictates which data is available for predictive modeling and how target outcomes are defined \cite{shroff2017predictive, saxena2020human}. This dimension is of critical importance since it directly impacts both human discretion and algorithmic decision-making. For instance, the policy decision to expand mandated reporting in child-welfare significantly increased the number of cases referred to CWS as well as broadened the definitions of child abuse and neglect (with implications for algorithmic modeling) \cite{melton2005mandated}. Prior work has acknowledged the dominant role that bureaucracy or policy plays in the public sector both in regard to decision-making as well as the adoption of technology \cite{lodato2018institutional, jackson2014policy}. The central role of bureaucratic processes is evident from prior work conducted in public services where caseworkers pushed for algorithmic systems that could help mitigate organizational contradictions and clarify organizational processes \cite{holten2020shifting}. 

\subsection{Algorithmic Decision-Making} 
Algorithmic Decision-Making is defined through the lens of \textit{street-level algorithms}, a term recently coined by Alkhatib and Bernstein \cite{alkhatib2019street} in the HCI community. Street-level algorithms directly interact with and make on-the-ground decisions about human lives and welfare in a sociotechnical system \cite{alkhatib2019street}. Prior work has argued that algorithms in the public sector is a domain in its own right \cite{holten2020shifting, de2020case, veale2018fairness, saxena2020human, saxena2020conducting} and must be characterized differently as compared to algorithms in the private sector where the decision outcomes are well-defined. Therefore, it becomes important to examine and critique the dimensions within Algorithmic Decision-Making that impact the predicted outcomes. Algorithmic Decision-Making is the most flexible element of the framework that designers can directly impact. That is, algorithms must be developed in such a way that they balance the other two elements (human discretion and bureaucratic processes). HCI methodologies such as value-sensitive algorithm design \cite{zhu2018value} and human-centered algorithm design \cite{baumer2017toward} can ensure that the algorithms account for values of stakeholders as well as theory-driven practice. Moreover, participatory design can unravel the policy mandates and institutional processes that often mediate the decision-making process and must be accounted for \cite{lodato2018institutional, saxena2020conducting}. 

\subsubsection{Relevant Data} 
Necessary information about individuals and their environment must be collected to be able to adequately predict an outcome of interest. In several domains within the public sector, there is significant debate about which predictors are associated with which outcomes \cite{saxena2020child, drawve2019utilizing, holten2020shifting}. Moreover, the necessary information may not always be available or inconsistently available with contradicting factors \cite{connelly2016role}. For instance, risk assessment algorithms in child-welfare have traditionally only used a narrow set of predictors (child and parent characteristics) to assess risk \cite{saxena2020human}. However, a more comprehensive understanding of risk is necessary, including the risk posed by the system itself \cite{gambrill2000risk}. Therefore, algorithms need to be theoretically constructed with proper considerations from domain experts with respect to feature selection and modeling to ensure that the algorithm offers higher utility and complements theory of practice \cite{saxena2020conducting, baumer2017toward, zhu2018value}.

\subsubsection{Types of Decision-Support} 
Two types of algorithms are predominantly used in the public sector; predictive and prescriptive algorithms. Predictive algorithms seek to predict the likelihood of the occurrence of an outcome of interest, whereas, prescriptive algorithms act as decision aids and offer recommendations to intervene and affect the outcome of interest \cite{deka2014big}. Examining the nature of decision-support systems is equally as important as interrogating the outcome itself. Prescriptive decision-aids are often introduced as a means to improve decision-making while not shifting agency away from workers. However, prior research shows that workers allow decision-aids to supplant their own decisions when they lack confidence and/or experience \cite{eubanks2018automating, saxena2020human, shlonsky2005next}. Moreover, the calls for human-in-the-loop might be moot if there is a lack of understanding about how algorithms impact human decision-making and how the type of decision-support (i.e.- algorithm design) impacts the practical possibilities for human intervention \cite{peeters2020agency, shlonsky2005next, schwalbe2004re}.

\subsubsection{Degrees of Uncertainty} 
Decision outcomes in the public sector are not well-defined and as previously noted, a person's life can stabilize or de-stabilize several times making it hard to predict what constitutes a successful outcome or intervention \cite{holten2020shifting}. Prior research has also established that an irreducible degree of uncertainty exists with respect to the outcomes in the public sector \cite{hammond1996human, de2020case, bullock2019artificial} with both humans and algorithms likely to make mistakes. Pääkkönen et al. \cite{paakkonen2020bureaucracy} further extend this argument to state that the design of algorithmic systems must identify and cultivate important sources of uncertainty because it is at these sources where the need for human discretion accumulates since ambiguity about the operation of the algorithm persists.

\vspace{0.1cm}
This framework challenges designers, practitioners, and policymakers to rethink the core assumptions and nature of their practice which are evolving in an increasingly socio-technical public sector and need to be re-examined in light of these new challenges and opportunities. It provides a structured way to think about socio-technical problems centered in algorithmic decision-making in the public sector, study the interdependencies between the dimensions, and recognize underlying causes that impact decision-making. 





\section{Methods}
In this section, we describe our partnership with a child-welfare agency to address the research questions set forth in the introduction of our paper.

\subsection{Study Overview}
The goal of this study was to examine the algorithms that caseworkers use in their daily work lives and unpack the collaborative nature of how these algorithms were used in group settings. To accomplish this, we partnered with a child-welfare agency, which serves about 900 families and 1300 children in a large metropolitan area in the midwestern United States. The state's Department of Children and Families (DCF) contracts child-welfare services to
this agency who must comply with all DCF standards including the use of mandated algorithms or decision-tools. We conducted an eight-month long in-depth ethnographic case study at the agency from August 2019 to March 2020. Before conducting observations or recruiting participants for interviews, we obtained Institutional Review Board (IRB) approval at our mid-sized private research university to conduct our study. We then emailed the participants an IRB approved consent form and obtained their verbal consent to participate in the study. During this time, the first-author observed child-welfare team meetings and conducted semi-structured interviews with key stakeholders at the agency. 

\subsection{Meeting Observations and Interviews}
The first author conducted in-person observations of meetings to gain the necessary understanding of how algorithms were used in a team setting and how caseworkers interacted with these algorithms in their daily work practices. These observations were also helpful in understanding the collaborative work of child-welfare teams that make decisions that are mediated by policies, social-work practice, and algorithms. Next, we provide a detailed description of the team meetings and interviews.

\subsubsection{45-Day Staff Meetings or Planning Meetings}
The 45-day staff meetings occur within the first 45 days of a case coming into the care of the agency and are attended by child-welfare team members involved at the front-end of case planning \cite{saxena2020child}. These meetings facilitate information sharing so that consensus can be reached in regard to the child's well-being and placement stability\footnote{Placement stability is defined as three or fewer placement moves for a foster child during the previous 36 months.}. Each meeting is scheduled for 90 minutes, and we observed 15 meetings. These meetings are typically attended by the CWS employees that work in case management, permanency planning, family preservation, and licensing. Central to these meetings is the \textit{7ei} staffing protocol that helps the child-welfare team apply principles and practices derived from trauma-informed care (TIC) \cite{hendricks2011creating} to each case. The 7ei Staffing protocol is accompanied by the 7ei algorithm which acts as a framework for TIC and helps track progress with respect to each case \cite{topitzes2019trauma}. The \textit{CANS} algorithm is also used at these meetings to establish a baseline for foster children with respect to mental health well-being. We also identified two more algorithms being used by the child-welfare teams. \textit{Legal Permanency Status (LPS)} algorithm is used to assess the quality of the current placement and recognize systemic barriers. \textit{Anti Sex-Trafficking (AST)} algorithm is used to assess the risk of sex-trafficking for foster youth. Observing these meetings allowed us to understand how these decision-making tools were being used in practice, the benefits they offered, as well as the challenges they posed.

\subsubsection{Permanency Consultation Meetings}
Permanency consultation meetings are specialized meetings designed to expedite permanency\footnote{Permanency is defined as reunification with biological parents, adoption or legal guardianship.} for children placed in out-of-home care by employing innovative best practices and seeking to address any systemic or policy-related barriers. These meetings are facilitated by permanency consultants and are staffed with many of the child-welfare team members that attend the planning meetings. They regularly occur at the 5, 10, and 15+ month marks for every case until the case is closed. These on-going meetings tended to be more involved than the planning meetings as limited information is available at the onset of a case. Moreover, permanency consultations involved cases that had been with the agency for several months (if not years) and revealed the messy interaction between the complex socio-political domain of child-welfare and the algorithmic tools being used. For instance, it was interesting to observe how the child-welfare teams reached consensus during decision-making discussions when they had to account for policy and systemic barriers, social-work practice, and the algorithms. Each meeting was scheduled for an hour, and we observed 40 of these meetings.


\subsubsection{Semi-Structured Interviews}
Next, we used the knowledge gathered from these observations to develop our interview protocol and recruit participants who consistently attended these meetings as part of their job routines. After having first observed the child-welfare teams interact with algorithms for several months, we conducted interviews to delve deeper into the participants' understanding of these decision-tools as well as the benefits and challenges as perceived by them. We asked participants a series of questions about the nature of child-welfare work and the algorithms we observed being used at the agency. We also asked them to expand upon any interactions we had observed during the meetings. For example, we asked them to share their appreciation or dissatisfaction towards a certain algorithm or feature, as well as their team's or self frustration with the misuse of these decision tools. We conducted 20 semi-structured interviews with child-welfare staff members, which included permanency consultants, supervisors, program directors, ongoing case managers, data specialists, and clinical therapists. Seventeen interviews were conducted at the child-welfare agency in the participants' private offices or conference rooms. Due to the COVID outbreak, the last three interviews were conducted over the phone to ensure the safety of both the participants and the researcher. Interviews lasted for about 45 minutes to 1 hour in duration. 

\subsection{Qualitative Data Analysis} 
The first author took detailed observational notes during each team meeting and compiled a debriefing document with his initial insights within 30 minutes of each meeting to retain as much of his thoughts as possible. The interviews were audio-recorded and transcribed verbatim for analysis. Notes, documents, and transcripts were shared among all co-authors. Our high-level research questions guided our analyses, but within those questions, we allowed for new insights to emerge and adjusted our research questions based on emergent insights. We performed a thematic qualitative analyses \cite{clarke2015thematic} to answer RQ1 and RQ3. The first author read through the interview transcripts several times and created initial codes and consulted with co-authors to form a consensus around the codes, as well as resolve any ambiguous codes. Next, these codes were conceptually grouped into themes. However, in our results, we also use our observational notes to augment the insights we gained from the interviews and note potential discrepancies and nuances from the holistic insights gained from our site observations.

\vspace{0.1cm}
For RQ1, we used an open-coding process to identify the high-stakes decision outcomes associated with the four algorithms that are embedded in child-welfare practice, namely, Child and Adolescent Needs and Strengths (CANS), Seven Essential Ingredients (7ei), Anti Sex-Trafficking Response Tool (AST), and Legal Permanency Status Tool (LPS). In understanding how these algorithms were used in the daily practices of the CWS employees, we identified seven key purposes (As shown in Table \ref{RQ1:codebook}): 1) \textbf{Compensation Calculations:} determine the monetary value to be offered to foster parents for caring for foster children, 2) \textbf{Mental Health Assessment:} conduct a mental-health screen of foster children to assess the risks and needs, 3) \textbf{Level of Foster Care:} determine a suitable placement setting capable of meeting the needs of children, 4) \textbf{Trauma-informed Care:} a trauma-responsive service model developed through an ecological understanding of adverse events and trauma experienced by children and families, 5) \textbf{Placement Stability and Permanency:} Track outcomes from the trauma-responsive service model and assess if they are leading to better outcomes, 6) \textbf{Sex Trafficking Risk Assessment:} assess the risk of sex-trafficking for a foster child, and 7) \textbf{Quality of Placements and Systemic Barriers:} Track the current quality of placement and the systemic barriers inhibiting permanency.

For RQ2, we used the ADMAPS framework to code for how each algorithm (i.e., CANS, 7ei, AST, LPS) impacted (positively, negatively, or both) dimensions of ADMAPS. For example, we found that CANS had an overall negative impact on human discretion with 80\% of the interviewees indicating that it negatively impacted professional expertise by limiting the scope for value judgments and heuristic decision-making; 75\% of interviewees saying that it reduced their ability to make flexible value judgments on child outcomes; and 80\% of interviewees asserting that they were no longer given the discretion to make decisions on behalf of the children assigned to them because CANS made several of these decisions for them. These mappings to the ADMAPS framework allowed us to assess the role that human discretion, algorithmic decision-making, and bureaucratic processes played with respect to each of the four algorithms deployed in CWS daily practice, as well as compare the differences between them. 

For RQ3, we synthesized the patterns across the four algorithms to identify emergent themes that were consistent across our analysis of RQ2 to give a bigger picture of the potential benefits and drawbacks associated when balancing the tradeoffs between human discretion, algorithmic decision-making, and bureaucratic processes.
 
\section{RESULTS}
In the following sections, we organize and present the results by our three research questions. First, we identify the high-stakes outcomes for which algorithmic decision-making is leveraged within CWS (RQ1) and the roles that human discretion, bureaucratic processes, and algorithmic decision-making play in these decision outcomes (RQ2). We do this separately for CANS, 7ei, AST, and LPS. Next, we discuss the potential benefits and drawbacks when balancing the different dimensions of ADMAPS framework (RQ3). Interviewees profiles can be found in the appendix.   

\subsection{{Child and Adolescent Needs and Strengths (CANS) Algorithm}}
The Child and Adolescent Needs and Strengths (CANS) algorithm was used at all the planning meetings (n=15) and discussed by all the interviewees (n=20). CANS algorithm is constructed using the CANS assessment; a communimetric tool designed to assess the level of need of a foster child and develop an individualized service plan \cite{lyons2004measurement}. With its primary purpose being communication, CANS assessment is designed based on communication theory rather than psychometric theories centered in measurement development \cite{lyons2009communimetrics}. CANS assessment was designed to support decision-making with respect to assessing a child’s level of need and service planning. However, as depicted in Figure \ref{fig:CANSoutcomes}, CANS algorithm has been re-appropriated to measure additional outcomes discussed below. It is conducted within the first 30 days of a child entering the child-welfare system or moving to a new placement (for e.g., foster home). It is then periodically conducted every six months.


\subsubsection{CANS High-Stakes Decision Outcomes (RQ1)} \leavevmode

\vspace{0.1cm}
\textbf{Mental Health Assessment and Level of Foster Care.}
All child-welfare workers at the agency are certified in conducting CANS for mental health screenings. As such, several participants (60\%, n=12) shared that the agency uses CANS algorithm to conduct a mental health assessment of foster children based on risks posed and exhibited behaviors. Participants (60\%, n=12) stated that the first CANS assessment was most useful because it helped establish a baseline for the mental-health needs of a child and the level of care that the child needed. For instance, one supervisor explained:

\begin{myquote}
    \textcolor{mygray}{\textit{"We need to have starting point.. what kind of behavioral issues does a child have? What are their needs? Because we can’t naively place a high-needs kid with foster parents who are not trained and certified to manage those needs. It’s a recipe for disaster"} -P9, Child Welfare Supervisor, MSW, 13 years}
\end{myquote}

The first assessment is used to devise service plans for foster children (for e.g, behavioral therapy) based on exhibited behaviors. It also facilitates sharing this information with other parties such as legal parties and family preservation specialists who also play a role in case planning. Child-welfare teams in both the planning meetings (n=15) and permanency consultations (n=40) briefly discussed child needs and behaviors but then shifted to a more extensive conversation about trauma using trauma-informed care. Some participants (50\%, n=10) also shared that CANS algorithm recommends the level of foster care that the child should be placed in (see Figure \ref{fig:CANSoutcomes}). Higher level foster homes are trained and certified in taking care of high-needs children. However, due to a lack of such homes, this decision often comes down to the availability of resources. 

\vspace{0.1cm}
\textbf{Compensation Calculations.} Using the CANS algorithm for calculating foster parents’ compensation was another prominent theme that emerged in 73\% (n=11) of the planning meetings and 85\% (n=17) of the interviews. However, it was not a dominant theme at the permanency consultations, because compensation is directly negotiated between the foster parents, case managers, and the supervisor. It does not require the input of rest of the child-welfare team. The state reimburses foster parents for the costs associated with having foster children placed in their homes. Most of the participants (85\%, n=17) quickly recognized CANS as the ``rate-setting tool'' even though compensation calculation was not the primary purpose of the algorithm. One supervisor explained:

\begin{figure}[t!]
  \includegraphics[scale=0.22]{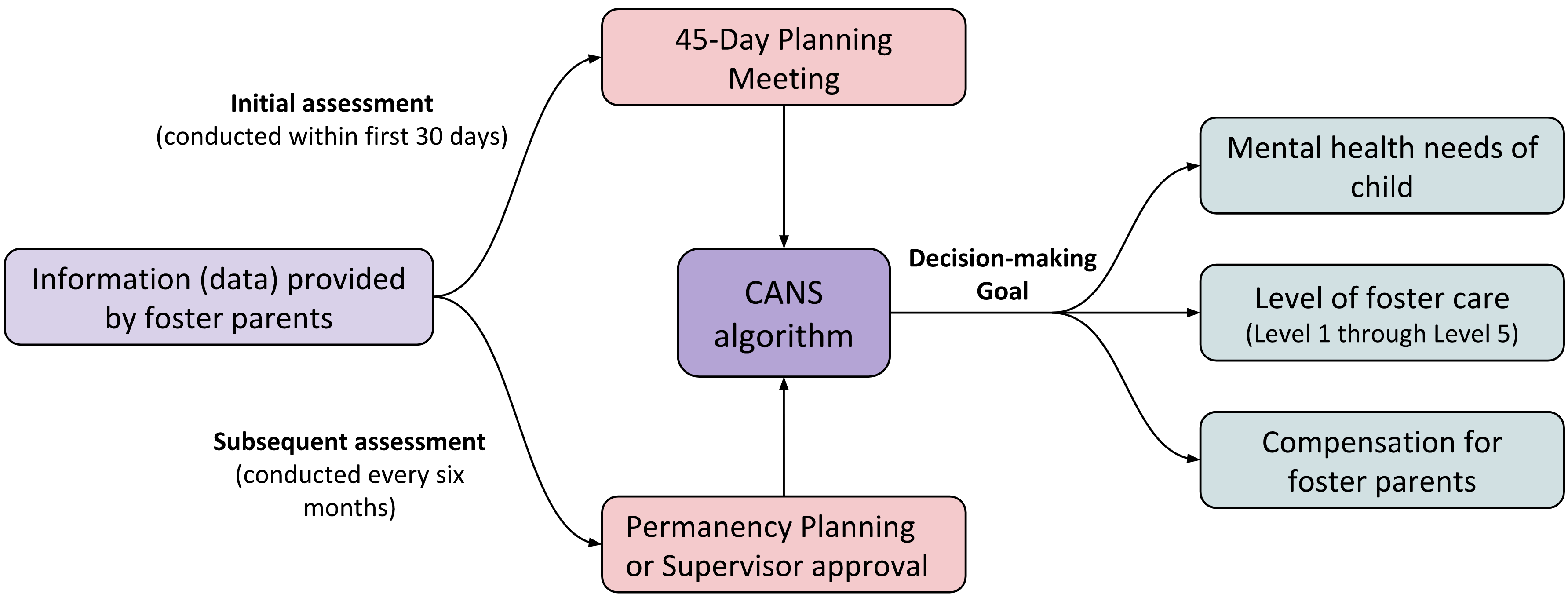}
  \caption{CANS Algorithm: Associated Decision Outcomes}
  \vspace{-0.2cm}
  \label{fig:CANSoutcomes}
  \vspace{-0.2cm}
\end{figure}

\begin{myquote}
    \textcolor{mygray}{\textit{"Foster parents who are taking in high-needs kids should be compensated for that. They have to put in significantly more time and energy in managing those behaviors, taking kids to therapy, setting healthy boundaries.. So we need this [CANS] standard to do that"} - P12, Child Welfare Supervisor, MSW, 7 years}
\end{myquote}

The state needs a \textit{metric} to be able to determine the compensation to be offered to each foster parent. The Department of Children and Families (DCF) decided to associate this compensation with the mental health needs of a child, that is, the higher the mental needs of a child, the higher the compensation offered to foster parents. Thus, even though the primary purpose of CANS was for conducting mental health assessments and for the level of foster care, it was re-appropriated to also calculate compensation associated with caring for a foster child's mental health needs. While it logically follows that a child with more mental health issues will require a higher level of care (consequently, higher cost of care), CWS employees were also well-aware that such cost-benefit analyses tied directly to something as subjective as mental health assessments were problematic.

\subsubsection{CANS and Algorithmic Decision-Making (RQ2)} 
Overall, caseworkers were frustrated that CANS misses important context about the child but is still used in a mandated predictive capacity. In the sections below, we discuss how CANS algorithm maps onto the dimensions of algorithmic decision-making (i.e., relevant data, type of decision-support, and degree of uncertainty). We use a percentage combined with an up or down arrow to denote the percentage of participants who indicated a positive or negative impact on each dimension of the ADMAPS framework for algorithmic decision-making, human discretion, and bureaucratic processes, respectively. We follow this structure throughout the remainder of our results. 

\vspace{0.1cm}
70\% \textcolor{red}{\faArrowDown} \textbf{Relevant Data.} \textit{CANS data does not account for trauma or social interactions}. Most of the participants (70\%, n=14) shared that CANS conducted the child’s assessment in an isolated manner and did not account for the quality and impact of relationships in their lives which are often more important for determining the long-term well being of these children. CANS algorithm focuses on the child emotional/behavioral needs (for e.g., anxiety, anger control, substance use, behavioral regression) and child risk behaviors (for e.g., suicide risk, self-harm, delinquent behavior, runaway tendencies) to assess the mental health needs of the child. For instance, a supervisor shared –

\begin{myquote}
    \textcolor{mygray}{\textit{"How do you measure empathy of others? You cant. Some foster parents are more empathetic and understanding and stand by the child no matter what. Sometimes that’s all it takes. You can’t put that in CANS"} -P15, Child Welfare Supervisor, 9 years}
\end{myquote}

Moreover, participants noted that CANS focuses on current behaviors but not the underlying trauma or traumatic triggers. The assessment is conducted based on exhibited behaviors over the past 30 days, however, the participants explained that trauma can stay with a child for years and lead to serious emotional dysregulation from time to time.

\vspace{0.1cm}
90\% \textcolor{red}{\faArrowDown} \textbf{Type of Decision-Support.}  \textit{The predictive nature of CANS leaves no room for discretion}. This challenge emerged in 90\% (n=18) of the interviews and 73\% (n=11) of the planning meetings. CANS algorithm is designed to predict or measure outcomes of interest once the data has been provided. Participants shared that the predictive nature of CANS leaves little room to exercise discretion and has become a great source of frustration for them. Gaming the inputs to achieve the desired outcome is the only way through which caseworkers regain agency. Participants explained that the higher the mental health needs of a child per CANS, higher the compensation offered to foster parents. After CANS scores are entered in, the algorithm generates a monetary value to be offered to foster parents. However, if foster parents disagree with the rate and believe they should be paid more, the caseworkers manipulate the scores to produce a higher rate. Most of the participants (90\%, n=18) shared their frustration regarding the inflexible and predictive nature of CANS. For instance, one supervisor explained:

\begin{myquote}
    \textcolor{mygray}{\textit{"Case managers and even supervisors are being forced to.. and kind of pressured into scoring children higher in order to provide higher numbers. So foster parents get paid more. CANS is a manipulative tool being used to barter off children...Children are being exploited for payment"} -P10, Child Welfare Supervisor, MS, LPC, NCC, 9 years}
\end{myquote}

\vspace{0.1cm}
70\% \textcolor{red}{\faArrowDown} \textbf{Degree of Uncertainty.} \textit{High degree of uncertainty associated with the outcomes}. Several participants (70\%, n=14) shared that CANS does not account for much of the data that they consider pertinent when assessing cases (for e.g., understanding about trauma and social support-system). This lack of relevant data leads to a high degree of uncertainty which is further exacerbated by the predictive and inflexible nature of CANS. Many felt that gaming the algorithm was the only viable option for caseworkers to exercise discretion. For instance, one supervisor shared:

\begin{myquote}
    \textcolor{mygray}{\textit{"CANS has become all about the rate. Generating the right rate so foster parents are happy with little to no attention paid to mental health needs"} -P13, Child Welfare Supervisor, MSW}
\end{myquote}


\subsubsection{CANS Severely Impedes Human Discretion (RQ2).} In this section, we discuss how CANS algorithm maps onto the ADMAPS dimensions of human discretion. 

\vspace{0.1cm}
80\% \textcolor{red}{\faArrowDown} \textbf{Professional Expertise.} \textit{CANS contradicts professional expertise}. All caseworkers at the agency are required to pass the CANS certification to be able to conduct the assessment with clients. However, most participants (80\% n=18) shared that their training in trauma-informed care (see Section 6.3), which offers a more comprehensive understanding of trauma and the child’s environment often conflicted with CANS. CANS further inhibited professional expertise because caseworkers felt that it turned them into data brokers who must collect information about children and feed it to the algorithm to make the decisions. The lack of relevant data and lack of decision-making latitude on part of caseworkers has turned CANS into primarily a rate-setting tool to calculate compensations. For instance, one case manager shared:

\begin{myquote}
    \textcolor{mygray}{\textit{"Caseworkers are doing CANS just to get it done.. to produce a good rate and reduce the conflict with foster parents. There is little to no attention paid to the mental health needs of kids."} -P17, Ongoing Case Manager, 8 years}
\end{myquote}

75\% \textcolor{red}{\faArrowDown} \textbf{Value Judgments.} \textit{CANS has introduced conflicting values}. Most of the participants (75\%, n=17) shared how the re-appropriation of CANS to calculate the foster parent compensations has led to several unintended consequences. CANS is re-conducted every six months to reassess the mental health needs of children, and consequentially, compensation is recalculated. However, with a focus on exhibited behaviors and not the underlying trauma, foster parents who are helping children cope and recover can see their compensation being lowered. A supervisor explained:

\begin{myquote}
    \textcolor{mygray}{\textit{"It is the complete opposite of what we want it [CANS] to do. Foster parents help minimize the behaviors and offer support so that kids can develop good coping skills. They help address the mental health needs and help kids stabilize by taking them to therapy and all their activities. But then they’re punished because the kid’s needs go down, and so does the rate"} -P9, Child Welfare Supervisor, MSW, 13 years}
\end{myquote}

Here, caseworkers are unable to prioritize properly conducting CANS because generating the adequate rate takes precedence to ensure the placement is not disrupted. Placement disruptions adversely affect foster children who develop emotional and behavioral problems and are unable to form lasting meaningful relationships with foster parents \cite{blakey2012review}. Therefore, caseworkers must prioritize supporting the current placement by any means necessary. This contradictory nature of CANS has turned caseworkers into value mediators and has left them frustrated because they are unable to adequately balance the needs of families and the demands of policymakers.

\vspace{0.2cm}
80\% \textcolor{red}{\faArrowDown} \textbf{Heuristic Decision-Making.} \textit{CANS leaves no room for heuristic judgment calls}. This theme emerged in 80\% (n=16) of the interviews and 73\% (n=11) of the planning meetings. With a lack of relevant information pertinent to a case and high degree of uncertainty, it becomes imperative that caseworkers are able to turn towards heuristics and make decisions with the assistance of the child-welfare team. However, CANS does not allow for heuristic decision-making with respect to the outcomes. Participants (80\%, n=16) shared that every case is contextually different with salient factors that might be central to one case but peripheral to another. This was also apparent in all the planning meetings and permanency consultations where the teams adaptively focused on information pertinent to that case.

\vspace{0.1cm}
\subsubsection{CANS and Bureaucratic Processes (RQ2).} Overall, CANS allocates resources but does not account for organizational constraints. In this section, we discuss how CANS algorithm maps onto the ADMAPS dimensions of bureaucratic processes.

\vspace{0.1cm}
40\% \textcolor{mygreen}{\faArrowUp} 70\% \textcolor{red}{\faArrowDown} \textbf{Resources \& Constraints.} \textit{CANS has introduced new constraints in the case planning process}. There is both a benefit and drawback to how CANS algorithm accounts for resources and constraints. Some participants (40\%, n=8) found value in the first assessment since it established the baseline for mental health and compensation for foster parents. Ideally, the algorithm offers an efficient way to allocate funds to foster parents based on mental health needs of the child. However, it does not account for organizational constraints and its implementation ends up introducing more constraints that frustrate caseworkers. Participants (70\%, n=14) shared that properly conducting CANS requires the caseworkers to interview several people, however, they are only able to interview foster parents due to high caseloads. A permanency consultant asserted:

\begin{myquote}
    \textcolor{mygray}{\textit{"You are supposed to interview foster parents and teachers as well as others that kids interact with to get a good CANS assessment. But with high caseloads, caseworkers only talk to the foster parents"} -P4, Permanency Consultant, MSW, APSW, 8 years}
\end{myquote}

However, as previously discussed, CANS is conducted every six months and foster parents are motivated to exaggerate behaviors to continue to be paid consistently. This is a constraint introduced by the implementation of the algorithm itself. Participants (50\%, n=10) also shared that the algorithm recommends the level of foster care that a child should be placed in, however, there is a lack of good foster homes in the system and this requirement is seldom met. Moreover, two data specialists shared that the caseworkers’ yearly job performance is tied to the timeliness with which they complete and submit CANS assessments since funds need to be allocated in a timely manner. Therefore, it exacerbates the need to get the assessments completed irrespective of the mental health needs of a child, and consequentially, adds another organizational constraint. 

\vspace{0.1cm}
40\% \textcolor{mygreen}{\faArrowUp} 80\% \textcolor{red}{\faArrowDown} \textbf{Administration \& Training.} \textit{Caseworkers are trained to conduct CANS but not on managing constraints}. This theme emerged in 80\% (n=16) of the interviews. CANS offers a way to support an organizational process of allocating resources. Caseworkers are trained and certified to conduct the mental health assessment, and as previously noted, some participants (40\%, n=8) found value in the first CANS assessment as a means to establish a baseline for mental health needs. However, they are not trained on how to manage the systemic constraints and value conflicts introduced by the algorithm itself. Participants (50\%, n=10) shared that new caseworkers socially learn from other caseworkers (and through experience) to manage these conflicts by gaming CANS. 

\vspace{0.1cm}
80\% \textcolor{mygreen}{\faArrowUp} \textbf{Laws \& Policies.} \textit{The use of CANS is legally mandated}. Most participants (80\%, n=18) recognized (and often complained) that CANS was legally mandated by the state and offered them a convenient means to allocate resources to foster parents every six months. Participants recognized that they must comply with this policy and continue implementing CANS enough though they routinely manipulated it to generate higher compensations.

\subsection{Seven Essential Ingredients (7ei) Algorithm}
\begin{figure}[t!]
  \includegraphics[scale=0.25]{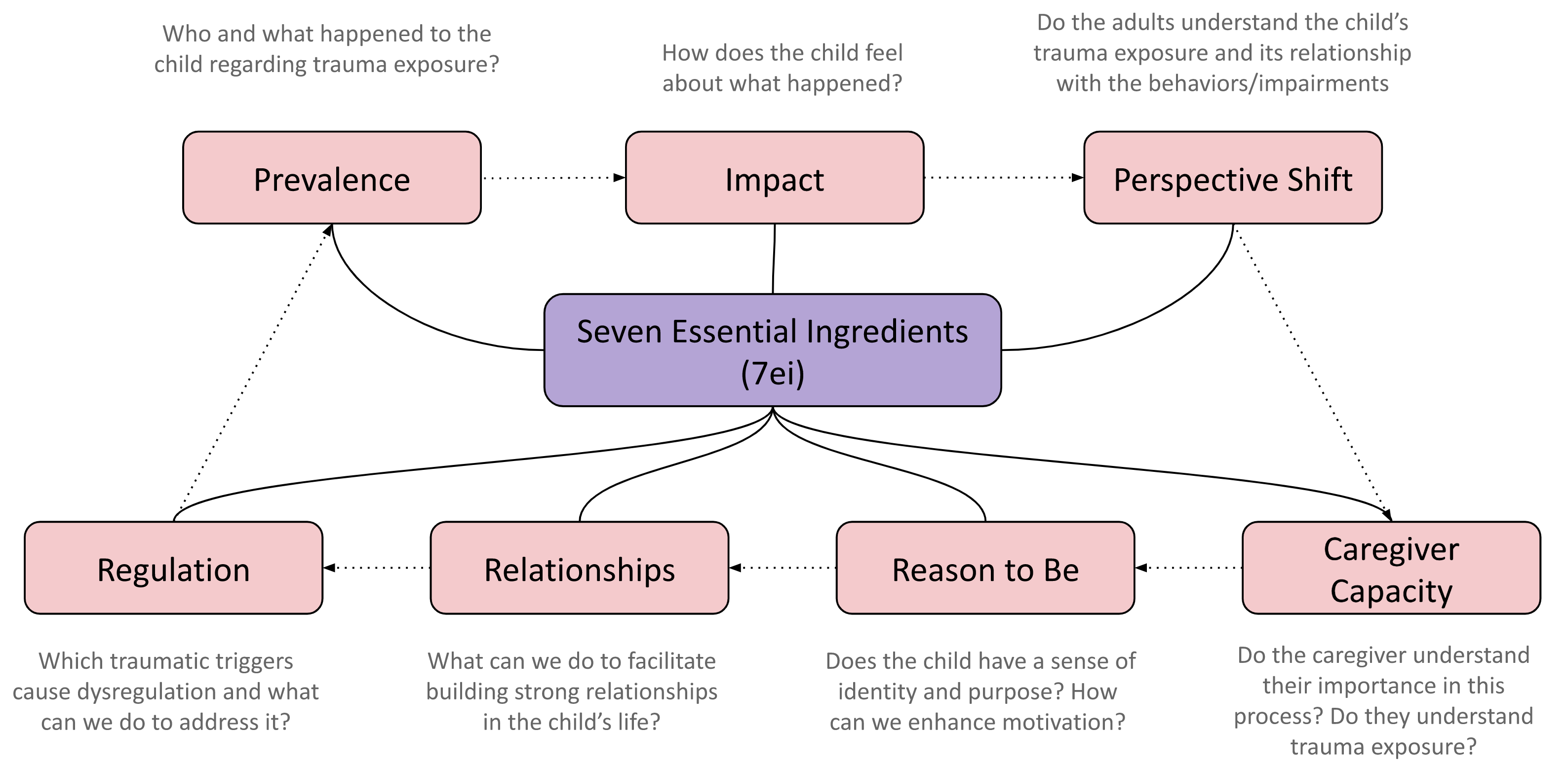}
  \vspace{-0.6cm}
  \caption{7ei Algorithm: Associated Decision Outcomes}
  \vspace{-0.3cm}
  \label{fig:7eioutcomes}
\end{figure}

The agency uses trauma-informed care \cite{hendricks2011creating} as their core guiding principle, which is embedded in their practice and trauma-responsive service model \cite{topitzes2019trauma}. 7ei is a theoretically constructed algorithm that is centered in the medical practice of trauma-informed care (TIC) and complements the agency's social work practice \cite{hendricks2011creating, k2010shelter}. We witnessed 7ei being utilized in all meetings (N=55). This algorithm complements specialized trainings developed by the agency for child-welfare workers and introduces the complexity of trauma, frameworks for understanding the effects of trauma, and the practices and principles of TIC \cite{topitzes2019trauma}. 7ei acts as a guiding framework and is used to track and score each case from a TIC perspective. The algorithm is designed to be used in a team setting such that team members can offer their expertise, reach consensus decisions, and devise case plans. The team discusses and scores the child’s and caregiver’s wellness on seven domains as depicted in Figure \ref{fig:7eioutcomes} and brainstorms solutions on how to make progress on these domains. 7ei algorithm based in TIC has also been proven to improve child outcomes such as placement stability and permanence \cite{topitzes2019trauma}. 7ei assessments are individually focused, however, unlike CANS, the results are trauma-focused and also guide family interventions, highlighting areas of child functioning upon which the caregivers and professionals should focus their attention \cite{topitzes2019trauma}.

\subsubsection{7ei High-Stakes Decision Outcomes (RQ1)}
The seven domains of 7ei are: \textit{Prevalence}, \textit{Impact}, \textit{Perspective Shift}, \textit{Regulation}, \textit{Relationships}, \textit{Reasons to Be}, and \textit{Caregiver Capacity}. As such, 7ei is not directly tied to a specific outcome such that every time 7ei is used an outcome is recommended or predicted. Instead of predicting an outcome of interest using other factors, 7ei is used to track outcomes (i.e.- 7ei domains) over time to assess the trajectory of a child-welfare case. It is primarily used as a prescriptive tool that serves as a framework for team-based brainstorming of solutions guided by a TIC framework. As such, the seven domains of 7ei are both the input variables and the output variables of the 7ei assessment process because these are both the considerations and outcomes that the team is trying to measure and improve over time. Improvement in the 7ei domains is associated with placement stability and permanency outcomes, however, the agency leadership resisted developing a singular aggregate index that would measure this outcome. Ongoing conversations with agency leadership (program directors, quality improvement leaders) revealed that assessments are more likely to be manipulated if they are tied to singular metrics. Agency leadership also uses 7ei to assess progress within the agency with respect to trauma-informed care. For instance, it helps them understand whether conversations and meetings founded in TIC are leading to better permanency outcomes. Case-level outcomes of 7ei that are discussed at all the planning meetings and permanency consultations are depicted in Figure \ref{fig:7eioutcomes}.

\subsubsection{7ei and Algorithmic Decision-Making (RQ2)}
We found that the prescriptive nature of 7ei helps adaptively select data and address uncertainties. In this section, we discuss how 7ei algorithm maps onto the ADMAPS dimensions of algorithmic decision-making.

\vspace{0.1cm}
90\% \textcolor{mygreen}{\faArrowUp} \textbf{Relevant Data.} \textit{7ei provides a comprehensive view of the child and their ecosystem}. 
Participants shared that 7ei algorithm offers a comprehensive view of the foster child, caregivers, impact of traumatic events in their life, as well as interactions in their social ecosystem. Moreover, the child-welfare teams in all the permanency consultations (n=40) and planning meetings (n=15) were able to adaptively select factors that were most pertinent to that case. Most participants (90\%, n=18) emphasized that each case carried a lot of nuance and could not be addressed based on a few broad set of predictors so they appreciated that 7ei allowed them to focus on certain factors and then brainstorm ideas on how to help the family make progress. For instance, a program director shared:

\begin{myquote}
    \textcolor{mygray}{\textit{“We have tried the cookie cutter approach in the past. Assigning everyone to parenting classes, therapy, and other family support services. It failed and it is horrible to do to a family. So, with 7ei we focus on addressing core issues whether it’s the parent’s self-esteem, their own abandonment issues or child’s emotional regulation that will really help this family” } -P7, Child Welfare Program Director, MSW, 20 years}
\end{myquote}

Timeline of the case established which 7ei domains the child-welfare teams converged on. The teams focused more on some domains than others based on how long the child had been in the system. For instance, in case of children who had been in the system for a few months, the team spent more time on \textit{Prevalence} and \textit{Impact} and then focused on \textit{Perspective Shift} (see Figure \ref{fig:7eioutcomes}). Trauma symptoms are associated with negative short-term outcomes such as placement instability \cite{clark2020investigating}, therefore, the team focuses on recognizing trauma early on so that proper interventions could be made that promote healing and improving outcomes. However, cases that had been in the system for a longer period or had experienced multiple placement moves, the team focused more on \textit{Regulation}, \textit{Perspective Shift}, and \textit{Reasons to Be}. Prior studies show that placements are often disrupted because foster parents are unprepared to manage behaviors of traumatized children \cite{carnochan2013achieving, taylor2014perspectives}, therefore, the team focuses on these domains to assess how to improve self-regulation for the foster child as well as expand the caregivers’ understanding of trauma.

\vspace{0.1cm}
85\% \textcolor{mygreen}{\faArrowUp} \textbf{Type of Decision-Support.}  \textit{The prescriptive nature of 7ei allows for brainstorming and idea generation}. Most participants (85\%, n=17) appreciated that 7ei allowed for open discussions and brainstorming of solutions. 7ei is designed to be used in a team setting and acts as the TIC framework such that the child-welfare team deliberates over each domain, scores it, and formulates an action plan through a trauma-informed perspective. A case manager shared: 

\begin{myquote}
    \textcolor{mygray}{\textit{"I like 7ei because we use it as a team, and it allows us to brainstorm ideas. We look at not just the behaviors but the underlying trauma that is causing those behaviors."} -P18, Case Manager, BSW, 2 years}
\end{myquote}

The 7ei domains are also the outcomes of interest that the team seeks to track and affect over time. Improvements in these domains are directly associated with the outcomes of placement stability and permanency which are tracked at the agency level.

\vspace{0.1cm}
80\% \textcolor{mygreen}{\faArrowUp} \textbf{Degree of Uncertainty.}  \textit{Adaptively selecting relevant data helps address high degree of uncertainty}. Most participants (80\%, n=16) felt that every case was contextually different and required their individual attention. That is, it was imperative they adequately weigh nuances and factors pertinent to that case. 7ei allows the child-welfare team to adaptively select the domains pertinent to a case and spend significantly more time on them and brainstorm solutions. Tracking outcomes over time and not using the tool in a predictive capacity allows the team to mitigate the high degree of uncertainty that would otherwise be associated with the predictions. 

\subsubsection{7ei Augments Human Discretion (RQ2).} In this section, we discuss how the 7ei algorithm maps onto the ADMAPS dimensions of human discretion.

\vspace{0.1cm}
85\% \textcolor{mygreen}{\faArrowUp} \textbf{Professional Expertise.} \textit{7ei helps develop professional expertise.} The algorithm is theoretically constructed and is centered in trauma-informed care. Participants (85\%, n=17) believed that it allowed them to brainstorm ideas from a TIC perspective and develop plans specific to that family. Continuous engagement with 7ei in trauma-informed meetings ensures that caseworkers are always thinking through TIC frameworks.  A permanency consultant asserted:

\begin{myquote}
    \textcolor{mygray}{\textit{"7ei isn’t just a "thing" that we do. It is centered in everything that we do. It ensures that caseworkers are always thinking through TIC frameworks"}  -P1, Permanency Consultation Supervisor, MSW, APSW, 22 years
}
\end{myquote}

This is especially important for CWS since the system lacks experienced caseworkers due to high turnover \cite{carnochan2013achieving}. Continually working through 7ei under proper supervision ensures that new caseworkers are developing professional expertise.

\vspace{0.1cm}
70\% \textcolor{mygreen}{\faArrowUp} \textbf{Value Judgments.} \textit{7ei allows caseworkers to make informed value judgments}. This theme emerged in 70\% (n=14) of the interviews and 72\% (n=40) of the meetings. 7ei is centered in social work’s core values of service, dignity, and worth of the person and allows caseworkers to prioritize these values and devise interventions that will directly help a child and their family cope with trauma. For instance, prioritizing the well-being of a child does not only mean sending them to therapy. It also incorporates addressing concerns within their ecosystem. A supervisor explained: 

\begin{myquote}
    \textcolor{mygray}{\textit{"Therapy only goes so far if nothing changes in the child’s ecosystem and they feel continually triggered by others. With 7ei we try to address problems in this ecosystem and devise approaches that will help the family"}  -P9, Child-Welfare Supervisor, 13 years
}
\end{myquote}


7ei allows the child-welfare team to make value-based judgments and take steps that improve the quality of a child's relationships. This takes the form of family-level interventions or sharing information with caregivers about the impact of trauma to bring about a perspective shift.

\vspace{0.1cm}
85\% \textcolor{mygreen}{\faArrowUp} \textbf{Heuristic Decision-Making.} \textit{7ei enables heuristic judgment calls}. Most participants (85\%, n=17) appreciated that 7ei offered them flexibility and autonomy in how they interact with it. This theme also emerged in all the planning meeting (n=15) and permanency consultations (n=40) where the child-welfare team adaptively selected information that was most pertinent to the case and often acted as an obstacle towards achieving permanency. For instance, cases where the child had experienced multiple placement moves, the team focused on \textit{Regulation} and \textit{Reasons to Be} to devise plans that would help improve the child’s emotional, behavioral, and cognitive functioning. 


\subsubsection{7ei Supports Bureaucratic Processes (RQ2).} In this section, we discuss how the 7ei algorithm maps onto the ADMAPS dimensions of bureaucratic processes.

\vspace{0.1cm}
60\% \textcolor{mygreen}{\faArrowUp} \textbf{Resources \& Constraints.} \textit{7ei accounts for the resources at the organization}. Participants (60\%, n=12) shared that 7ei is locally developed at the agency and accounts for the resources available at the agency in the form of supervision, expertise, and specialized trainings. Participants also shared that critical decision-making power in regard to achieving permanency sits with the legal parties (district attorneys, judges), however, 7ei operates within these constraints and allows caseworkers to help families and prepare them to be able to receive a favorable decision in court.

\vspace{0.1cm}
75\%\textcolor{mygreen}{\faArrowUp} 55\%\textcolor{red}{\faArrowDown}  \textbf{Administration \& Training.}  \textit{7ei is embedded in daily work processes but requires additional oversight and training}. This theme emerged in 75 \% (n=15) of the interviews and all the planning meetings (n=15) and permanency consultations (n=40) and there is both a benefit and a drawback associated with it. Participants (75\%, n=15) recognized that continually engaging with the tool in meetings was helpful but some participants (55\%, n=11) shared that the tool also added more tasks to an already heavy workload. The agency offers specialized trainings for trauma-informed care with 7ei acting as a tool that complements these trainings. Moreover, we observed that 7ei is embedded into daily work processes because it is utilized every time a case is discussed in a team setting. This continued engagement under supervision has helped earn the trust of caseworkers who learn not just from the algorithm but also the collective expertise of the child-welfare team. For instance, a case manager shared:

\begin{myquote}
    \textcolor{mygray}{\textit{"I like 7ei because we use it as a team, and it allows us to brainstorm ideas. It just helps.. thinking out loud with everyone and knowing that I don't have make these decisions alone. Also, it helps guide my thought process but doesn’t tell me what to do."} -P17, Case Manager, BSW, 8 years}
\end{myquote}

However, 7ei has added to the workload of both the case managers and supervisors who must discuss and complete an additional tool as part of their job requirement at the agency. 

\vspace{0.1cm}
65\% \textcolor{red}{\faArrowDown} \textbf{Laws \& Policies.} \textit{7ei is not legally mandated and only used locally at the agency}. Several participants (65\%, n=13) were frustrated by the fact that 7ei was not legally mandated and that they had to continue using CANS. Even though an independent research study showed that 7ei is leading to better permanency and placement outcomes, the algorithm still lacks legitimacy at the state and federal level. 


\subsection{Anti Sex-Trafficking (AST) Algorithm}
An emerging high-stakes decision for which an algorithm is used is assessing the risk of sex-trafficking for a foster child over 10 years of age. We observed this in 65\% (n=13) of the interviews, 33\% (n=5) of the planning meetings, and 35\% (n=14) of the permanency consultations. The agency has a dedicated team called HART (Human Anti-Trafficking Response Team) that manages cases where the foster child might be at high risk of being trafficked. If a child meets the criteria for risk of sex-trafficking per the AST algorithm, the case must be reported to HART. Early identification of such indicators can play a significant role in ensuring child safety.



\subsubsection{AST High-Stakes Decision Outcomes (RQ1)} Risk indicators are divided into three domains - "At-Risk", "High-Risk", and "Confirmed". The child-welfare team must continue to closely monitor the case if they select fewer than three "At-Risk" indicators and continue to have conversations to mitigate the risks. If the team selects three or more "At-Risk" indicators or 1 or more "High-Risk" or "Confirmed" indicators, then the case must be referred to HART.

\subsubsection{AST and Algorithmic Decision-Making (RQ2)}
Overall, we found that AST algorithm often missed important context, which frustrated caseworkers. In this section, we discuss how AST algorithm maps onto the ADMAPS dimensions of algorithmic decision-making.

\vspace{0.1cm}
65\% \textcolor{mygreen}{\faArrowUp} 75\% \textcolor{red}{\faArrowDown} \textbf{Relevant Data.} \textit{AST assesses pertinent risk indicators but misses context about the case}. Several participants (65\%, n=13) shared that the algorithm offers a new perspective on sex-trafficking by capturing risk indicators that they had not considered before. For instance, “possession of money, electronics, cosmetics, or clothes that are unexplained”, “traveling out of the area or somewhere out of the ordinary”, and “unwilling to provide information about an older partner” are some of these indicators. A supervisor shared:

\begin{myquote}
    \textcolor{mygray}{\textit{"It’s helpful to have us think about it differently. I have been doing this for a very long time and when I started, we were never thinking about trafficking. So now we are more conscious of these risks" } -P8, Child Welfare Supervisor, MSW, APSW, 19 years}
\end{myquote}

However, participants (75\%, n=15) also shared that the algorithm misses context about each case and the presence of some of these risk indicators did not mean the child was being sex trafficked.

\vspace{0.1cm}
75\% \textcolor{red}{\faArrowDown} \textbf{Type of Decision-Support.} \textit{Predictive nature of AST frustrates caseworkers}. Several participants (75\%) who found value in the algorithm as a guide also shared their frustrations in regard to its mandatory reporting nature. AST is used in a predictive capacity such that if certain risk indicators are selected then the case must be reported to HART.  For instance, a permanency consultant shared:

\begin{myquote}
    \textcolor{mygray}{\textit{"If a child is at risk for sex trafficking then we are having those conversations from the very beginning and taking necessary action. This decision tool is not helpful in the way it’s being used and only frustrates HART"} -P5, Permanency Consultant. MSW, 12 years}
\end{myquote}
 
Reporting to HART takes the case away from the child-welfare team who have spent significant amount of time building a relationship with the child and their caregivers. Moreover, HART is receiving an influx of calls that do not require their expertise as a result of this algorithm.

\vspace{0.1cm}
75\% \textcolor{red}{\faArrowDown} \textbf{Degree of Uncertainty.} \textit{Lack of context about the case leads to high degree of uncertainty}. Several participants (75\%, n=15) shared that even though AST was useful as a guide, the presence of risk indicators did not mean the child was at risk of being trafficked. There was still a lot of pertinent information that was necessary to make such a determination. For instance, a supervisor shared that one of their foster kids had a history of sexual abuse and met a bunch of criteria on the tool, however, the supervisor’s team is actively involved with the child and their caregivers, understand their needs, and did not believe the case needed to be reported to HART.  


\subsubsection{AST Impedes Human Discretion (RQ2)} In this section, we discuss how AST algorithm maps onto the ADMAPS dimensions of human discretion.

\vspace{0.1cm}
70\% \textcolor{mygreen}{\faArrowUp} \textbf{Professional Expertise.} \textit{AST has made caseworkers more aware of risk indicators}. Most participants (70\%, n=14) appreciated that this decision-tool taught new caseworkers to be actively aware of indicators that are associated with sex trafficking. As stated by P5 above, AST has improved professional expertise since caseworkers are more cognizant of such risks.

\vspace{0.1cm}
75\% \textcolor{red}{\faArrowDown} \textbf{Value Judgments and Heuristic Decision-Making.} \textit{AST does not allow caseworkers to make value judgments or engage in heuristic decision-making}. Several participants (75\%, n=15) shared that the risk indicators were just part of a bigger picture and there was still a lot context that the child-welfare team needed to collectively unpack to better assess the situation. This requires the child-welfare team to engage in heuristic decision-making and weigh the necessary information. For instance, a case manager shared:

\begin{myquote}
    \textcolor{mygray}{\textit{"One of my foster teens is sexually active and has a long-term boyfriend. But looking at the tool, everyone is like... ‘Is that her boyfriend or her pimp?’ I know this kid and she trusts me... building trust with these kids takes time... she doesn’t need to be referred to HART [Human Anti-Trafficking Response Team]"} -P17, Case Manager, BSW, 8 years}
\end{myquote}

Here, the case manager emphasizes that having trusting relationships was equally important for preventing foster youth from being sex trafficked. Referring the case to HART takes this case away from the case manager and the child is assigned a new case manager from HART.

\subsubsection{AST and Bureaucratic Processes (RQ2)} Overall, we found that AST is inadequately supported by bureaucratic processes which leads to frustrations on part of caseworkers. In this section, we discuss how AST algorithm maps onto the ADMAPS dimensions of bureaucratic processes.

\vspace{0.1cm}
60\% \textcolor{red}{\faArrowDown} \textbf{Resources \& Constraints.} \textit{AST has caused an influx of cases referred to HART}. Several participants (60\%, n=12) shared that AST has added to the frustrations of HART who do not have enough resources to manage all the cases that are being reported to them because of the mandatory reporting aspect of AST. This leaves HART with significantly less resources to focus on the cases that need them. A supervisor on HART explained:

\begin{myquote}
    \textcolor{mygray}{\textit{“It's become overused and is being abused. It's being used in different settings and not how it was originally intended to be used. We are getting an influx of calls that don't need to be called.”} -P10, Child Welfare Supervisor, MSW, LPC, NCC, 9 years}
\end{myquote}

\vspace{0.1cm}
65\% \textcolor{mygreen}{\faArrowUp} 60\% \textcolor{red}{\faArrowDown} \textbf{Administration \& Training.} \textit{AST is conducted as an organizational process and offers some training}. Most participants (65\%, n=13) shared that AST is routinely conducted for foster youth over 10 years of age. As discussed previously, it trains new caseworkers to be more conscious about risk factors but often to the detriment of HART. Participants also consider this decision-tool to be peripheral to the case planning process for most cases and utilized it as a mandated requirement.

\vspace{0.1cm}
60\% \textcolor{mygreen}{\faArrowUp} \textbf{Laws \& Policies.} \textit{The use of AST is mandated by law}. The Department of Children and Families (DCF) has legally mandated the use of this algorithm as a means to proactively protect foster youth as well as collect more data about associated risk factors and the number of foster youth considered at risk. Next, we discuss the Legal Permanency Status (LPS) algorithm.

\subsection{Legal Permanency Status (LPS) Algorithm} Tracking performance metrics such as placement stability and permanency emerged as another decision outcome for which an algorithm is being used. Several participants (65\%, n=13) discussed this decision-tool, and we observed the tool being used at all the permanency consultations (n=40). Federal legislation has established permanency as one of the primary goals of CWS and requires agencies to meet this well-defined and measurable benchmark \cite{williams2014child}. Legal permanency is defined as reunification with the biological family, adoption or transfer of guardianship. \cite{koh2008propensity}. 

\subsubsection{LPS High-Stakes Decision Outcomes (RQ1)}
This algorithm is used to track an outcome over time instead of predicting an outcome of interest using input data. The agency uses it at permanency consultations to track the quality of the current placement and if those specialized meetings are leading to better outcomes. The tool is facilitated by the permanency consultant and upon completion, the team categorically rates the quality of the placement. It is also used to track systemic barriers that are getting in the way of achieving permanency. Some participants recognized (40\%, n=8) the utility of this tool and a permanency consultant explained – 

\begin{myquote}
    \textcolor{mygray}{\textit{"It helps us to be actively aware of where we are at in terms of permanency. We have to follow a [15-month] timeline for permanency and if parents are not showing initiative and not completing court ordered services then we need to start exploring placement options. So, this tool is kind of an extra push"} -P5, Permanency Consultant, MSW, 12 years}
\end{myquote}
 

\subsubsection{LPS and Algorithmic Decision-Making (RQ2)} Overall, we found that even though LPS is used in a prescriptive capacity, it still lacks utility. In this section, we discuss how LPS algorithm maps onto the ADMAPS dimensions of algorithmic decision-making.

\vspace{0.1cm}
65\% \textcolor{red}{\faArrowDown} \textbf{Relevant Data.} \textit{Definitions of input variables are ambiguous and frustrate caseworkers}. Several participants (65\%, n=13) shared their indifference towards LPS and stated that the definitions of input variables in terms of what constitutes a “Good”, “Fair”, or “Poor” placement were ambiguous. A permanency consultant who is tasked with conducting this decision-tool shared:

\begin{myquote}
    \textcolor{mygray}{\textit{“This tool is confusing in itself; the definitions are very vague. We can't necessarily put families into categories. We have to use this tool and have conversations about where they [families] might fit best, but it doesn’t give a clear picture of what's really going on with the placement.”} - P6, Permanency Consultant, BSW, 3 years}
\end{myquote}

Participants (65\%, n=13) also shared that how LPS was conducted depended on the perspective of the team. Different people might look at the same set of facts and reach different conclusions as to why the placement should be rated as good, fair, or poor. 

\vspace{0.1cm}
60\% \textcolor{red}{\faArrowDown} \textbf{Type of Decision-Support.} \textit{LPS is used as a prescriptive tool, but caseworkers lack agency towards affecting outcomes}. Participants (60\%, n=12) shared that the tool did not predict or recommend an outcome based on the inputs, however, the utility of the tool lies in recognizing the current state of a case with respect to permanency and addressing systemic barriers, but the child-welfare team lacks agency with respect to addressing several systemic barriers. For instance, critical decision-making power in regard to permanency decisions sits with the legal parties (district attorneys and judges) who can choose to fully disregard the child-welfare team’s recommendation. 

\vspace{0.1cm}
65\% \textcolor{red}{\faArrowDown} \textbf{Degree of Uncertainty.} \textit{Lack of relevant information leads to high degree of uncertainty}. Participants (65\%, n=13) were most frustrated by the lack of relevant information that LPS needed to account for in order to make a proper determination about the quality of the placement. Moreover, there is high uncertainty associated with the timeliness with which some systemic barriers can be addressed making it hard to assess the quality of the placement (see P2 quote below).

\subsubsection{LPS impedes Human Discretion (RQ2)} In this section, we discuss how LPS algorithm maps onto the ADMAPS dimensions of human discretion.

\vspace{0.1cm}
40\% \textcolor{mygreen}{\faArrowUp} \textbf{Professional Expertise.} \textit{LPS builds professional expertise by establishing an urgency towards permanency}. Some participants (40\%, n=8) shared that the decision-tool was useful in that it established a sense of urgency and insistence towards achieving permanency for foster children. It teaches new caseworkers that they needed to prioritize finding placement options even if that upset biological parents. The child-welfare team follows a 15-month timeline where the parents must complete court order services within this timeline to achieve reunification. Towards the end of this timeline, CWS team must begin exploring alternate placement options.

\vspace{0.1cm}
60\% \textcolor{red}{\faArrowDown} \textbf{Value judgments and Heuristic Decision-Making.} \textit{LPS tool does not allow for value judgments or heuristics}. Participants (60\%, n=12) shared that the tool does not allow for value judgments or heuristics on part of the child-welfare team since decision-making power about permanency decisions sits with the legal parties.

\subsubsection{LPS and Bureaucratic Processes (RQ2)} Overall, LPS supports bureaucratic processes but does not account for organizational constraints. In this section, we discuss how LPS algorithm maps onto the ADMAPS dimensions of bureaucratic processes.

\vspace{0.1cm}
65\% \textcolor{red}{\faArrowDown} \textbf{Resources \& Constraints.} \textit{LPS does not account for organizational resources and imposes new constraints}. Participants shared (65\%, n=13) that ambiguity around input variables and lack of agency with respect to permanency decisions has turned this decision-tool into documentation that every child-welfare team must complete at permanency consultations. Moreover, there are arbitrary constraints placed on the decision-tool in regard to how a placement can be rated. For instance, a program director explained:

\begin{myquote}
    \textcolor{mygray}{\textit{"We may have a court hearing date set and all of a sudden, the placement is “Good”. We now have a "Good" rating because we have a court hearing. But it might take three years to terminate parental rights or go to a guardianship. And then you went three years without achieving permanency but somehow, we have a “Good" placement rating."} -  P2, Permanency Consultant, MSW, 20 years}
\end{myquote}

\vspace{0.1cm}
80\% \textcolor{red}{\faArrowDown} \textbf{Administration \& Training.} \textit{Caseworkers are indifferent towards how LPS is administered}. Even though the tool is expected to be central to permanency consultations, we noticed that for majority of these meetings (87\%, n=35), LPS was used towards the end as a requirement to rate the quality of the placement. Most participants (80\%, n=16) were indifferent towards LPS because of the several constraints and utility issues discussed above. One supervisor shared:

\begin{myquote}
    \textcolor{mygray}{\textit{"It’s just another thing we have to do in the permanency meetings. I let the Permanency Consultants score it however they like"} - P15, Child Welfare Supervisor, MSW, 9 years}
\end{myquote}

\vspace{0.1cm}
65\% \textcolor{mygreen}{\faArrowUp} \textbf{Laws \& Policies.} \textit{The use of LPS tool is legally mandated and helps track performance outcomes}. Participants (65\%, n=13) shared that the Department of Children and Families (DCF) has mandated using this decision-tool as a way to track the important outcomes of quality of placements and permanency. Moreover, it’s important to track the systemic barriers that often impact permanency to formulate policies at the state level that address these systemic barriers.




\subsection{Assessing the Benefits and Drawbacks of Differing Approaches (RQ3)}
 Table \ref{RQ3:arrowtable} offers a summary of how the four algorithms balance the dimensions of the framework. In this section, we discuss the benefits and drawbacks that arise when trying to balance the tradeoffs between human discretion, algorithmic decision-making, and bureaucratic processes.

\subsubsection{Algorithmic Decision-Making Should Seek to Augment Human Discretion, Not Supplant it.} In this section, we discuss the themes around the benefits and drawbacks that emerged when balancing the ADMAPS dimensions of human discretion and algorithmic decision-making.

\vspace{0.2cm}
\textbf{When aligned, algorithms augment decision-making processes, but a lack of alignment can take away autonomy and heuristic decision-making.} All of our participants emphasized that every case is contextually different and that a family’s circumstances often change through the life of the case. For instance, a parent seeking reunification with their child might experience a lapse while trying to maintain a stable job, maintain their sobriety, or consistently attend court ordered services. In other words, algorithms should be designed with the recognition that there will be a high degree of uncertainty associated with any relevant data and the subsequently predicted outcome. Therefore, algorithms need to not only make room for human discretion but also facilitate value judgments and heuristics in order to offer utility. Most of our participants (80\%, n=16) mentioned that 7ei algorithm augmented their decision-making processes when they were making difficult decisions. It prioritizes and enhances the value judgments and heuristic decision-making that caseworkers must engage in when devising action steps to help families. Participants (85\%, n=17) especially appreciated that 7ei facilitated brainstorming and idea generation instead of predicting an outcome of interest. One case manager explained:

\begin{myquote}
    \textcolor{mygray}{\textit{"Of all the things we have brought up, 7ei is my favorite because its helps us think differently, understand what a family has been through and then brainstorm ideas on how to help them based on this understanding of trauma"} -P3, Permanency Consultant, MSW, APSW, 9 years}
\end{myquote}


Participants noted that the tool offers flexibility and autonomy in how they interact with it and which 7ei domains they focused on. Adaptively selecting information instead of analyzing all the information is a key feature of heuristic decision-making and can lead to more accurate decisions. On the other hand, tensions arise when algorithms attempts to supplant human discretion. This is the case with CANS whose predictive nature does not account for the high degree of uncertainty that accompanies each case, and consequently, does not make room for discretionary work on part of the caseworkers. With a lack of autonomy, gaming the algorithm is the only way caseworkers are able to exercise discretion and produce the desired outcome. One supervisor shared: 

\begin{myquote}
    \textcolor{mygray}{\textit{"CANS is all about producing a good rate so foster parents can afford the resources they need to take care of the child. I have had foster parents put in notices [to end placement] because they couldn't support the child anymore"} -P14, Child Welfare Supervisor, MSW, 30 years}
\end{myquote}

This inflexible and predictive nature has shifted focus away from the primary outcome of interest (i.e. – mental health screening) and towards the secondary outcome that allocates resources.

\begin{table}[]
\Small
\begin{tabular}{@{}|l|ll|ll|ll|@{}}
\toprule
                       & \multicolumn{2}{c|}{\textbf{Human Discretion}}                                                                                                               & \multicolumn{2}{c|}{\textbf{Algorithmic Decision-Making}}                                                                                                                                                                     & \multicolumn{2}{c|}{\textbf{Bureaucratic Processes}}                                                                                                                                                                          \\ \midrule
                       & \cellcolor[HTML]{FFE0D2}Professional Expertise    & \cellcolor[HTML]{FFE0D2}80\% \textcolor{red}{\faArrowDown}   & \cellcolor[HTML]{FFE0D2}Relevant Data         & \cellcolor[HTML]{FFE0D2}70\% \textcolor{red}{\faArrowDown}                                                                        & \cellcolor[HTML]{FFECB3}Resources/Constraints & \cellcolor[HTML]{FFECB3}40\% \textcolor{mygreen}{\faArrowUp} 70\% \textcolor{red}{\faArrowDown} \\
                       & \cellcolor[HTML]{FFE0D2}Value Judgments           & \cellcolor[HTML]{FFE0D2}75\% \textcolor{red}{\faArrowDown}   & \cellcolor[HTML]{FFE0D2}Decision-Support      & \cellcolor[HTML]{FFE0D2}90\% \textcolor{red}{\faArrowDown}                                                                        & \cellcolor[HTML]{FFECB3}Admin \& Training     & \cellcolor[HTML]{FFECB3}40\% \textcolor{mygreen}{\faArrowUp} 80\% \textcolor{red}{\faArrowDown} \\
\multirow{-3}{*}{\textbf{CANS}} & \cellcolor[HTML]{FFE0D2}Heuristic Decisions & \cellcolor[HTML]{FFE0D2}80\% \textcolor{red}{\faArrowDown}   & \cellcolor[HTML]{FFE0D2}Uncertainty & \cellcolor[HTML]{FFE0D2}70\% \textcolor{red}{\faArrowDown}                                                                        & \cellcolor[HTML]{DEF3D0}Laws \& Policies      & \cellcolor[HTML]{DEF3D0}80\% \textcolor{mygreen}{\faArrowUp}                                                                      \\
\hline
                       & \cellcolor[HTML]{DEF3D0}Professional Expertise    & \cellcolor[HTML]{DEF3D0}85\% \textcolor{mygreen}{\faArrowUp} & \cellcolor[HTML]{DEF3D0}Relevant Data         & \cellcolor[HTML]{DEF3D0}90\% \textcolor{mygreen}{\faArrowUp}                                                                      & \cellcolor[HTML]{DEF3D0}Resources/Constraints & \cellcolor[HTML]{DEF3D0}60\% \textcolor{mygreen}{\faArrowUp}                                                                      \\
                       & \cellcolor[HTML]{DEF3D0}Value Judgments           & \cellcolor[HTML]{DEF3D0}70\% \textcolor{mygreen}{\faArrowUp} & \cellcolor[HTML]{DEF3D0}Decision-Support      & \cellcolor[HTML]{DEF3D0}85\% \textcolor{mygreen}{\faArrowUp}                                                                      & \cellcolor[HTML]{FFECB3}Admin \& Training     & \cellcolor[HTML]{FFECB3}75\% \textcolor{mygreen}{\faArrowUp} 55\% \textcolor{red}{\faArrowDown} \\
\multirow{-3}{*}{\textbf{7ei}}  & \cellcolor[HTML]{DEF3D0}Heuristic Decisions & \cellcolor[HTML]{DEF3D0}85\% \textcolor{mygreen}{\faArrowUp} & \cellcolor[HTML]{DEF3D0}Uncertainty & \cellcolor[HTML]{DEF3D0}80\% \textcolor{mygreen}{\faArrowUp}                                                                      & \cellcolor[HTML]{FFE0D2}Laws \& Policies      & \cellcolor[HTML]{FFE0D2}65\% \textcolor{red}{\faArrowDown}                                                                        \\
\hline
                       & \cellcolor[HTML]{DEF3D0}Professional Expertise    & \cellcolor[HTML]{DEF3D0}70\% \textcolor{mygreen}{\faArrowUp} & \cellcolor[HTML]{FFECB3}Relevant Data         & \cellcolor[HTML]{FFECB3}65\% \textcolor{mygreen}{\faArrowUp} 75\% \textcolor{red}{\faArrowDown} & \cellcolor[HTML]{FFE0D2}Resources/Constraints & \cellcolor[HTML]{FFE0D2}60\% \textcolor{red}{\faArrowDown}                                                                        \\
                       & \cellcolor[HTML]{FFE0D2}Value Judgments           & \cellcolor[HTML]{FFE0D2}75\% \textcolor{red}{\faArrowDown}   & \cellcolor[HTML]{FFE0D2}Decision-Support      & \cellcolor[HTML]{FFE0D2}75\% \textcolor{red}{\faArrowDown}                                                                        & \cellcolor[HTML]{FFECB3}Admin \& Training     & \cellcolor[HTML]{FFECB3}65\% \textcolor{mygreen}{\faArrowUp} 60\% \textcolor{red}{\faArrowDown} \\
\multirow{-3}{*}{\textbf{AST}}  & \cellcolor[HTML]{FFE0D2}Heuristic Decisions & \cellcolor[HTML]{FFE0D2}75\% \textcolor{red}{\faArrowDown}   & \cellcolor[HTML]{FFE0D2}Uncertainty & \cellcolor[HTML]{FFE0D2}75\% \textcolor{red}{\faArrowDown}                                                                        & \cellcolor[HTML]{DEF3D0}Laws \& Policies      & \cellcolor[HTML]{DEF3D0}60\% \textcolor{mygreen}{\faArrowUp}                                                                      \\
\hline
                       & \cellcolor[HTML]{DEF3D0}Professional Expertise    & \cellcolor[HTML]{DEF3D0}40\% \textcolor{mygreen}{\faArrowUp} & \cellcolor[HTML]{FFE0D2}Relevant Data         & \cellcolor[HTML]{FFE0D2}65\% \textcolor{red}{\faArrowDown}                                                                        & \cellcolor[HTML]{FFE0D2}Resources/Constraints & \cellcolor[HTML]{FFE0D2}65\% \textcolor{red}{\faArrowDown}                                                                        \\
                       & \cellcolor[HTML]{FFE0D2}Value Judgments           & \cellcolor[HTML]{FFE0D2}60\% \textcolor{red}{\faArrowDown}   & \cellcolor[HTML]{FFE0D2}Decision-Support      & \cellcolor[HTML]{FFE0D2}60\% \textcolor{red}{\faArrowDown}                                                                         & \cellcolor[HTML]{FFE0D2}Admin \& Training     & \cellcolor[HTML]{FFE0D2}80\% \textcolor{red}{\faArrowDown}                                                                        \\
\multirow{-3}{*}{\textbf{LPS}}  & \cellcolor[HTML]{FFE0D2}Heuristic Decisions & \cellcolor[HTML]{FFE0D2}60\% \textcolor{red}{\faArrowDown}   & \cellcolor[HTML]{FFE0D2}Uncertainty & \cellcolor[HTML]{FFE0D2}65\% \textcolor{red}{\faArrowDown}                                                                        & \cellcolor[HTML]{DEF3D0}Laws \& Policies      & \cellcolor[HTML]{DEF3D0}65\% \textcolor{mygreen}{\faArrowUp}                                                                      \\ \bottomrule
\end{tabular}
\vspace{0.1cm}
\caption{RQ3: Tradeoffs between balancing the dimensions within human discretion, algorithmic decision-making, and bureaucratic processes, Percentages (\%) represent the proportion of participants who stated that a dimension was positively (or negatively) impacted by the algorithm.} 
\vspace{-0.6cm}
\label{RQ3:arrowtable}
\end{table}

\vspace{0.2cm}
\textbf{When aligned, algorithms can help embed important value judgments into the decision-making process}. Most of the participants (80\%, n=16) felt strongly about the need to support each family in a different capacity and through different practices. This was also a dominant theme in all the planning meetings (n=15) and permanency consultations (n=40) where the child-welfare team devised specific plans for each family through trauma-informed care (i.e. - using 7ei). 7ei algorithm is centered in some of social work’s core values of service, dignity and worth of the person, and importance of human relationships \cite{topitzes2019trauma} and informs the child-welfare team’s work processes. For instance, a supervisor explained: 

\begin{myquote}
    \textcolor{mygray}{\textit{"7ei helps us think about how we can help every family. What can we do to help mom develop her self-worth? How can we help her build relationships with relatives or people in the community, so she has more caregiver support"} -P8, Child Welfare Supervisor, MSW, CAPSW, 19 years}
\end{myquote}

Several participants (65\%, n=13) explained that 7ei scores translated into actionable steps that directly sought to help children and their families. In the meetings, 7ei brainstorming sessions resulted in solutions that the child-welfare team could affect directly and not simply refer children and parents to therapy. For instance, the case managers and supervisors planned activities that the family could engage in to improve child as well as family functioning, discussed information to share with foster and biological parents about impacts of trauma exposure, as well as steps caregivers could take to establish healthy boundaries with foster children and enforce positive discipline. For instance, a permanency consultant explained:

\begin{myquote}
    \textcolor{mygray}{\textit{"It [7ei] helps you have a perspective shift on the family and the child. You don’t need to refer everyone and their mother to therapy. Sometimes it’s just as simple as having them do something as a family that’s different than what they ever did before [picnics, sports].. and challenge them in different ways"} -P2, Permanency Consultant, MSW, 20 years}
\end{myquote}

On the contrary, algorithms that do not embed human values into its design may consequentially end up minimizing them. For instance, CANS recommends the level of foster care that the child should be placed in and based on that the child-welfare team finds foster parents who have the resources to meet those needs. However, participants explained that more financial resources do not always equate to a foster child’s well-being. For instance, a permanency consultant explained: 

\begin{myquote}
    \textcolor{mygray}{\textit{"Everyone has a different compass for well-being. We had a child who was placed with well-off foster parents in a five-bedroom house and he wasn't doing well there because it was a culture shock for him. He wasn't used to a huge home, a great school, a big backyard... and he completely shut down. So, we found his aunt and we moved him down to Chicago. It's a two-bedroom house with five other kids and this child is thriving! So, he needed to be with his family, and he needed to have his own culture.} -P4, Permanency Consultant, MSW, APSW, 8 years}   
\end{myquote}

Here the permanency consultant emphasizes the core human value of having trusting relationships in one’s life and the importance of one’s culture that play a critical role towards achieve emotional and cognitive well-being. Unlike CANS, 7ei allowed the child-welfare team to prioritize \textit{Reasons To Be} and \textit{Regulation} to really help children instead of simply focusing on financial resources. Caseworkers must continually negotiate values and balance the individual needs of people with the demands of policymakers. Algorithms that do not allow caseworkers to make such value judgments may lead to more frustrations and limit their utility.


\vspace{0.1cm}
\textbf{Algorithms that replace the need for caseworker expertise can lead to inadequate and unreliable decision-making}. Another dominant theme that emerged in 70\% (n=14) of the interviews, 53\% (n=8) of the planning meetings, and 50\% (n=20) of the permanency consultations was that the algorithms in use were diminishing the need for caseworker expertise and family-centered care in a very contextual domain. CWS is a high-stakes domain where most caseworkers lack adequate work experience, carry high caseloads, and are under a lot of pressure from legal parties. Most of the participants (70\%, n=14) expressed concerns that algorithmic systems might simply act as a safe default for most caseworkers such that questioning an algorithmic decision would add more work to their plate. A program director explained – 

\begin{myquote}
    \textcolor{mygray}{\textit{"We are not hiring people with a lot of experience. All new hires are recent graduates who don’t quite know what this field looks like. They are happy to trust the machine to just get through the day"} -P7, Child Welfare Program Director, MSW, 20 years}
\end{myquote}

Professional expertise in the public sector deteriorates when algorithms limit the scope for value judgments and heuristic decision-making. Value negotiations and heuristics are an indispensable aspect of professional practice that workers must continually engage in to build expertise \cite{loyens2010toward}. There are also growing concerns that algorithms such as CANS are leading to children being referred to unnecessary services which also shifts the focus away from family-centered care. CANS scores directly translate to actionable steps in the form of services that children are referred to. Caseworkers co-opt CANS to produce higher compensations for foster parents or the foster parents might exaggerate child behaviors, however, several participants (60\%, n=12) recognized that this also meant unnecessary services being requested for children. One case manager explained: 

\begin{myquote}
    \textcolor{mygray}{\textit{"Foster parent might exaggerate behaviors just to get more money. And then we put the kid in therapy and are not addressing their needs specifically. The kid in therapy is then being asked `why are you so sad?’ and the kid is like.. `I’m not sad.’ "} -P17, Case Manager, BSW, 8 years}
\end{myquote}
 
This is further problematic because unnecessary services are not only an added financial burden on CWS but they also add to the medical trauma of foster children who are continually being told that something is wrong with them. Finally, as previously noted, services assigned through an isolated view of a child might be less effective than family-level interventions developed through a trauma-informed perspective. Algorithms like CANS, however, limit child-welfare workers from using their expertise and developing more family-centered practices.


\subsubsection{Algorithmic Decision-Support Systems and Bureaucratic Processes Need to be Aligned.} In this section, we discuss the themes around the benefits and drawbacks that emerged when balancing the ADMAPS dimensions of algorithmic decision-making and bureaucratic processes. 

\vspace{0.2cm}
\textbf{When algorithmic decision-making and bureaucratic processes are aligned, it can help train caseworkers.} Most participants (70\%, n=14) recognized the value of algorithms as essential training mechanisms. Continuous engagement with 7ei in trauma-informed meetings under proper supervision ensures that caseworkers are continually having conversations centered in trauma and are coming up with solutions founded in TIC. This is essential because CWS suffers from a high turnover with lack of adequate training and supervision being some of the leading causes \cite{carnochan2013achieving}. During the observations, experienced members of the child-welfare team such as the supervisors and permanency consultants brainstormed ideas with the caseworkers using 7ei and shared practices and approaches that had worked in the past with other families. A program director explained:

\begin{myquote}
    \textcolor{mygray}{\textit{"It [7ei] makes us think differently and in the moment think through TIC [trauma-informed care].. what is the impact of that [incident]? How can we help support parents and children and help with emotional regulation? What can we do to build their relationships and how do we support the people in their life so that they show better caregiver capacity to manage what this child is going through"} -P7, Child Welfare Program Director, MSW, 20 years}
\end{myquote}

7ei meetings are also attended by Caregiver Support Specialists and Family Preservation Specialists who based on case circumstances also offer their expertise to the caseworkers on how to proceed. Several participants (70\%, n=14) also recognized the Anti Sex-Trafficking algorithm as a good training tool such that caseworkers are always aware of and looking for red flags associated with sex-trafficking. This tool is also used in a team-setting and allows for the less experienced caseworkers to learn from the seasoned members of the team in how to perceive certain risks as well as how to follow up on them without confronting the foster youth. One supervisor explained: 

\begin{myquote}
    \textcolor{mygray}{\textit{"Its a whole change of mindset and and we’re now more cognizant of some of those red flags. So I appreciate that. So, anytime in supervision, if they [caseworkers] start talking about some of these things, we pull up the tool and go through it and start discussing how to proceed"} -P15, Child Welfare Supervisor, 9 years}
\end{myquote}

However, participants who found great value in the Anti Sex-Trafficking algorithm as a training mechanism were also equally frustrated by the mandatory reporting nature of it. 

\vspace{0.1cm}
\textbf{When the algorithmic decision-support system is not aligned with the constraints of the bureaucratic system, it leads to utility issues.} The utility of algorithmic decision-support systems is severely limited when they do not account for the organizational constraints within which they must operate. This challenge and frustrations on part of caseworkers was observed in 80\% (n=16) of the interviews, 73\% (n=11) of the planning meetings, and 82\% (n=32) of the permanency consultations. Most of the participants (80\%, n=16) stated that they were unable to use algorithms as intended because of several organizational constraints such as limited time availability due to high caseloads. For instance, properly conducting CANS requires the caseworkers to interviews several individuals such as the foster parents, relatives, and teachers to be able to get consistent information. However, most caseworkers have high caseloads and do not have enough time to devote towards properly conducting each assessment. A permanency consultant explained: 

\begin{myquote}
    \textcolor{mygray}{\textit{"It can take hours to properly do CANS. Everyone is stretched too thin.. we literally do not have the time to do that."} -P2, Permanency Consultant, MSW, 20 years}
\end{myquote}

Moreover, the bureaucratic policy that requires CANS to be conducted every six months makes it unfeasible for caseworkers to interview anyone beyond the foster parents. However, as previously noted, the assessment is also tied to the compensation offered to foster parents which further leads to foster parents exaggerating child behaviors. A supervisor explained that the algorithm did not need to be manipulated and caseworkers could request an exceptional rate to account for several different factors such as transportation services, therapy, school activities etc. However, she also explained that high caseloads make it hard to devote any extra time to each assessment:

\begin{myquote}
    \textcolor{mygray}{\textit{"The supervisor or worker can take a little extra time and put an exceptional amount in. But again, everybody has too much to get done and it’s unlikely for somebody to even think about that until they are in the middle of submitting CANS and its due the next day."} -P8, Child Welfare Supervisor, MSW, CAPSW, 19 years}
\end{myquote}

Caseworkers who continue to feel disempowered and frustrated by CANS find it easier to manipulate CANS scores to produce a higher rate than request the addition of an exceptional rate which needs to be approved by the supervisor. Most participants (75\%, n=15) also shared significant concerns about the fidelity of data being used to develop algorithms. Administrative data curated through bureaucratic processes can not be uncritically used to develop algorithms. Participants shared that the data about families is collected by caseworkers and it was hard to acquire accurate information. Oftentimes, caseworkers might receive conflicting information from parents, relatives, and neighbors. A supervisor explained:

\begin{myquote}
    \textcolor{mygray}{\textit{"People aren’t willing to happily share personal information about themselves. And understandably so. It’s really hard to be able to get consistent information to be able to put in some data system to bust out a decision. So I find this really difficult to comprehend how we would even consider that"} - -P13, Child Welfare Supervisor, MSW, 12 years}
\end{myquote}

Families view caseworkers as representatives of the state and are unwilling to trust them or share any information that they might consider incriminating. Therefore, much of the information is based on the caseworker's perception of the family. This problem is further exacerbated by the fact that most caseworkers are new graduates and lack adequate experience. A program director explained that during investigations, caseworkers need to ask the right questions, read situations, and follow-up to be able to derive meaningful information. However, caseworkers acquire these skills through years of experience.


\vspace{0.1cm}
\textbf{If the algorithms are not explained well to the practitioners, they don’t trust them.} Most of the participants (85\%, n=17) were unaware of the algorithms being used. They did not recognize CANS or 7ei as algorithmic systems, even though they interact with them on a daily basis. This finding is especially interesting because the CANS training material recognizes it as an algorithm and explicitly lays out the purposes (i.e. - outcomes) that the algorithm is designed to accomplish. CWS started using algorithms as a means to track important performance metrics such as permanency and placement stability as mandated by federal legislation \cite{wilson2014integrating}. Moreover, several psychometric assessments that are routinely used in child-welfare to assess the risks and needs of children and parents have adopted algorithmic analogues in the form of risk assessment algorithms. They are now being used as data collectors for all cases as well as decision arbiters for future cases. The CANS algorithm is a case in point of this scenario which is an algorithmic version of the CANS communimetric assessment \cite{lyons2009communimetrics} and has been re-purposed for other outcomes. Interestingly, caseworkers were well aware of assessments used at the agency but did not actively recognize the algorithmic components unless the interviewer nudged them to think about some of the automated aspects. For instance, when explicitly discussing CANS, all the interviewees realized it to be an algorithm that plays a pertinent role in decision-making. We asked a permanency consultant how she thought CANS worked and she exclaimed: 

\begin{myquote}
    \textcolor{mygray}{\textit{"Yes, okay! Right! Because we put in the scores, and it generates the rate. Oh and I hate the CANS! I, I’ve expressed it a lot of times that CANS should not be tied to money."} -P6, Permanency Consultant, BSW, 3 years}
\end{myquote}

As soon as participants recognized the automated decision-making aspects of these decision tools, they started to share some other ambiguous aspects. For instance, they had trouble recognizing who managed these decision-tools and shared that different tasks associated with these tools were distributed across the agency. Most of the participants (85\%, n=17) where unable to isolate CANS to a particular role or "place". CANS is embedded in several aspects of the case planning process with different departments involved. Case management team focuses on assessing risks and needs whereas the data specialists track scores, and the fiscal liaisons review and approve foster parent compensations. Participants (70\%, n=14) also alluded to the fact that algorithms meant different things to people based on their intended goals. For instance, one permanency consultant asserted:

\begin{myquote}
    \textcolor{mygray}{\textit{"Case management try to do a good job with risks and needs but of-course foster parents only care about the rate. I know X’s team [data specialists] look at scores to see if there is improvement. But there is also Y [fiscal liaison] who just wants the scores to be turned in on time so that he doesn’t have to chase people down"} -P1, Permanency Consultation Supervisor, MSW, APSW, 22 years}
\end{myquote}
 
Several participants (75\%, n=15) recognized that CANS, AST, and LPS decision tools meant different things in different contexts. Lack of awareness about algorithms at the organizational level and their distributed nature augmented caseworkers’ distrust.


\subsubsection{Collective Buy-in Requires Trust in Outcomes at Both the Caseworker and Bureaucratic Levels.} In this section, we discuss the themes around the benefits and drawbacks that emerged when balancing the ADMAPS dimension of human discretion and bureaucratic processes with respect to decision outcomes that the CWS agency implements.

\vspace{0.2cm}
\textbf{When the algorithmic outcomes miss important context, practitioners don’t trust them} This challenge emerged in 70\% (n=14) of the interviews, 67\% (n=10) of the planning meetings and 75\% (n=30) of the permanency consultations. Algorithms attempt to generalize child and family characteristics and place them in certain categories to be able to make a determination. However, this inadvertently leads to a loss of information with respect to the final outcome since all the information cannot be accounted for by these algorithms. For instance, the CANS algorithmic assessment focuses on child behaviors, risks, and needs based on the last 30 days and does not account for traumatic triggers that can lead to serious emotional dysregulation from time to time. A permanency consultant explained:

\begin{myquote}
    \textcolor{mygray}{\textit{"CANS does not account for trauma in the way 7ei does. We have a child that goes into manic depression every year around holiday season and needs to be cared for 24/7. So, one of the foster parents has to quit their job. However, there is no way to account for that in CANS. This child is doing fine right now but we know that traumatic trigger is coming."} -P3, Permanency Consultant, MSW, APSW, 9 years}
\end{myquote}

Participants shared similar concerns about the Anti Sex-Trafficking tool. Participants (65\%, n=13) shared that the tool was useful to identify risk factors, however, the presence of risk factors does not necessarily mean that the foster youth is being trafficked. A supervisor explained:

\begin{myquote}
    \textcolor{mygray}{\textit{"We have a 12 year old boy who suffers from a lot of sexual trauma. He met a bunch of criteria on the tool and we had to report to HART. We have been with this boy for many years.. we know him and what his behaviors are connected to. Reporting to HART felt very odd"} -P11, Child Welfare Supervisor, MSW, APSW, 9 years}
\end{myquote}

Here, the supervisor alludes to the fact that presence of risk factors did not capture the full picture and missed important context about what is going on with this child. As previously discussed, child-welfare teams are having on-going conversations about the risk of sex-trafficking from the onset of a case and are able to investigate concerns without reporting to HART. Continued interactions illustrated by these examples leads to an accumulation of distrust towards algorithms that often miss important nuances and context.



\vspace{0.1cm}
\textbf{There is collective buy-in when practitioners believe that better outcomes are being achieved.}
This theme emerged in 85\% (n=17) of the interviews, 67\% (n=10) of the planning meetings, and 62.5\% (n=25) of the permanency consultations. Participants appreciated 7ei and trauma-informed care because it allowed them to directly help families by developing specific plans for them. Moreover, they were aware that trauma-responsive services developed through TIC lead to better permanency and placement outcomes for children. When frustrated by algorithms such as CANS, caseworkers often asked why they were still expected to use CANS when 7ei was leading to better outcomes. 

The agency developed a comprehensive four-part evidence-based service program centered in trauma-informed care of which the algorithmic tool is just one component. The first component of this program involves extensive ongoing TIC trainings. The second component introduced child-welfare staff to trauma-informed assessments that resulted in family-level interventions. The third component was the availability of specialized supervision and consultations provided by a clinical supervisor, TIC program administrator, medical expert, and a caregiver support specialist. Finally, the fourth component of this program is the 7ei algorithm which allows the child-welfare team to break down each case into seven domains and apply TIC principles and practices to it. In sum, the agency significantly invested in resources to ensure that TIC practice and 7ei were fully-supported by bureaucratic processes. A better understanding of what a family might be going through at a psychological level and knowing exactly how to help them has led to the collective buy-in from caseworkers. Most participants (85\%, n=17) found great value in TIC and appreciated 7ei as a tool designed to be a guiding framework for TIC. One permanency consultant explained:

\begin{myquote}
    \textcolor{mygray}{\textit{"Compassion fatigue in child welfare is very real. This tool has helped me truly understand what trauma can do to the brain. So I think now that we have a better understanding of trauma... we also have more empathy, and we go the extra mile to help families."} -P1, Permanency Consultation Supervisor, MSW, APSW, 22 years}
\end{myquote}
 
This was also evident at the meetings where the child-welfare staff urged case managers to ask and think in terms of \textit{"What happened to you and how can I help you?"} instead of \textit{"What is wrong with you?"} when working with children and families. That is, having a perspective shift and thinking in terms of the impact of trauma and not just exhibited behaviors. Moreover, a study published by independent researchers showed that child-welfare teams who implemented TIC practices using 7ei exhibited improved permanency and placement stability outcomes for their cases. This has further deepened the collective buy-in from caseworkers who often brought up the "study" to state that their practice centered in TIC worked and lead to better outcomes, and therefore, the legal parties needed to trust their judgments more. However, interestingly, none of the participants except the program director were able to locate the study and share it with us.


\vspace{0.1cm}
\textbf{Lack of trust in algorithmic outcomes, leads to concerns about unethical and unsound decision-making.}
This theme emerged in 80\% (n=16) of the interviews, 40\% (n=6) of the planning meetings, and 45\% (n=18) of the permanency consultations where participants recognized that algorithms in use sometimes led them to make unethical or unsound decisions. Surprisingly, even though 90\% (n=18) of the participants recognized that caseworkers were gaming the system to produce higher compensations, some participants (40\%, n=8) did not consider the decisions made to be unethical or unsound. To them, this is how the system was set up to be used. With contradictory incentive structures and conflicting values, caseworkers are often put in a position where they are forced to make such decisions. As previously discussed, both caseworkers and foster parents are co-opting CANS to produce a higher compensation, however, most participants (90\%, n=18) recognized that the base compensation offered to foster parents is pretty low and gaming the algorithm was the only convenient way to produce an adequate rate. A supervisor explained:

\begin{myquote}
    \textcolor{mygray}{\textit{ "There is a lack of good foster homes and there is no financial incentive to be doing this work... So caseworkers do whatever they can to get them [foster parents] the money they need or we risk disrupting a placement...Several foster parents have put in notices in the past because they can’t financially sustain the placement"} -P9, Child Welfare Supervisor, MSW, 13 years}
\end{myquote}



Another challenging aspect of using CANS where it leads to unsound decision-making is that it is now being used to track progress with respect to mental health; an unintended outcome of the continuous data collection. Consequently, an improvement (or deterioration) in mental health is impacting placement decisions. A clinical therapist explained: 

\begin{myquote}
    \textcolor{mygray}{\textit{"Sometimes a kid isn’t doing well because they are working through things. For example, when a kid starts therapy sometimes their behaviors gets worse, but it might actually be that they’re working through some things in therapy. And that’s not captured by the tool. When there is no context, I just can’t interpret whether or not that worsening is actually a bad thing."} -P16, Clinical Therapist, LCSW, 5 years}
\end{myquote}

These frustrations were also consistent with the fact that CANS is tied to service-planning where CANS recommends unnecessary child-level services when family-level interventions are often more effective in helping the family heal and cope with trauma. Prior studies have found that caseworkers distrust risk assessment algorithms such as CANS because of their deficit-based nature \cite{brown2019toward}. Interestingly, participants who were averse to this deficit-based nature did find value in risk assessment in regard to understanding a family's history. One permanency consultant explained:

\begin{myquote}
    \textcolor{mygray}{\textit{"The data gives you a bigger picture of what the family is going through. Because the families are not always honest with us about what they need. And so if you can see that they were in drug and alcohol services or housing authority... that would be helpful in child welfare. Not necessarily giving them a score. I don’t even know what to do with the score. It tells me nothing."} -P6, Permanency Consultant, BSW, 3 years}
\end{myquote}

Understanding a family’s past helps caseworkers assess the type of services that would be most beneficial for them. Looking at this data from a trauma-informed perspective allows caseworkers to better understand the underlying trauma in the family that might be leading to exhibited behaviors and subsequent interactions with the child-welfare system. This points to a change in mindset in regard to the intended goal of risk assessment of (i.e. - predicting the risk of future harm) towards a strength-based outcome of providing the right services to families and not necessarily labeling them as high or low risk. The permanency consultant here also invokes social work's core value of service and helping families and sees the utility in risk assessment through that perspective.


\section{Discussion}

Our results provide implications for algorithmic decision-making in CWS, more broadly for the public sector, as well as specific design guidelines for developing such systems in the public sector.
\vspace{-0.1cm}
\subsection{Implications for Algorithmic Decision-Making in the Child-Welfare System}
\subsubsection{Identifying Gaps and Opportunities for Algorithmic Decision-Making for Child-Welfare through the ADMAPS Framework}
Our results suggest that caseworkers have differing perceptions of the design, outcomes, and intended uses of the various algorithms. ADMAPS dimensions helped uncover caseworker perspectives with respect to different aspects of algorithms. For instance, examining the bureaucratic processes at the agency revealed that the most frustrating part about CANS is that it is being used to calculate financial resources for foster parents. Furthermore, deliberating over the relevant data and degree of uncertainty with the participants revealed that CANS did not account for traumatic triggers in a child’s ecosystem or the lack of  interpretation regarding worsening behaviors. That is, ADMAPS uncovered the multiple and conflicting roles of CANS as a compensation calculator as well as a mental health assessment tool. On the other hand, participants considered 7ei to be analogous to the trauma-informed care framework but not an algorithm that was continually collecting data and tracking outcomes over the life of a case. Differing socio-technical imaginations of algorithms in the public sector are aligned well with existing literature \cite{brown2019toward, saxena2020human, seaver2017algorithms, holten2020shifting, robertson2021modeling}. Moreover, while many of our participants were able to connect how CANS risk scores were used in making decisions about a child based on state guidelines, yet others were unable to make this connection since CANS is deeply embedded within their daily culture, where they did not recognize it as an algorithm anymore. This aligns with prior literature on thinking about algorithms as part of culture \cite{seaver2017algorithms} and points to a need in raising more awareness in caseworkers about the algorithms that they use daily. One way to improve this is to support explanations of algorithms in daily use; not just interpretations of outcomes but also designing around other interactions (e.g. data collection, input, visualization etc). Recent work on explainable AI has found support for focusing on explanations of daily interactions of users with algorithms \cite{liao2020questioning, muchatowards}. In addition, transparency of data, methods, and outcomes can support collaborative algorithmic decision-making processes \cite{schmidt2020calibrating, hois2019achieve}. Finally, assessing the scope of bureaucratic processes and how algorithms must function within these constraints allows for better practitioner engagement and leads to algorithms that are centered in practice.

\vspace{-0.1cm}
\subsubsection{Balancing Strength-based and Deficit-based Approaches}
Our results show that there are differing approaches in algorithmic implementation that is mandated by the state as opposed to the non-profit agency that is contracted by the state, to provide child-welfare services. For instance, the state mandates the use of CANS, which the caseworkers are not very amenable to as they can see the deficiencies in using this algorithm to provide good services for children in reality. In response, the agency developed 7ei using best practices from health services and social work (Trauma-Informed Care)  \cite{topitzes2019trauma, levenson2017trauma}, and continue to use this algorithm in parallel with CANS to provide a more holistic perspective for the best outcome for a child and family. As outlined in our results, CANS and 7ei affect the three dimensions of ADMAPS differently because fundamentally, CANS is a deficit-based assessment that focuses on risk mitigation and resource allocation; on the other hand, 7ei is a strength-based assessment that focuses on improving the outcomes for children. This is not to imply that CANS is always employed in a deficit-based context while 7ei only focuses on building strengths. These kinds of assessments can be commonly found in public services beyond child-welfare. For instance, within criminal justice systems, deficit-based risk assessments are the norm \cite{stevenson2018assessing} but are heavily criticized for being unfair \cite{berk2018fairness}. An important takeaway is that we need both strength and deficit-based approaches to make better decisions in high stakes environments. An initial assessment of risk is necessary but a transition towards helping people through strength-based approaches is equally important to prevent referrals and future interactions with the system \cite{badillo2017abandoned, badillo2018stakeholders, badillo2018chibest, badillo2017understanding}. Moreover, regardless of the type of outcome, street-level bureaucrats in the public sector must still exercise discretion and contextualize the algorithmic results for each case and act within the constraints posed by bureaucratic processes as explained by the ADMAPS framework \cite{alkhatib2019street, paakkonen2020bureaucracy, holten2020shifting}.

\vspace{-0.1cm}
\subsection{Implications for Algorithmic Decision-Making in the Public Sector}

\subsubsection{ADMAPS Emphasizes Managing Interdependencies and Trade-offs in Algorithmic Decision-Making Within the Public Sector}
The purpose of ADMAPS is twofold : 1) to interrogate algorithmic interventions in the public sector in a way that ensures that they balance the dimensions of \textit{human discretion} and \textit{bureaucratic process} with \textit{algorithmic decision-making}, and 2) to offer practical guidelines for developing algorithms that offer higher utility to practitioners. A key implication of using ADMAPS is that better algorithmic decisions are made when algorithms account for and balance the complex interdependencies within socio-technical systems, rather than operate in isolation. We found that when one aspect of the framework was optimized, the other aspects of the framework often suffered. For instance, CANS demonstrated an over-reliance on predictions to support mandated bureaucratic processes; therefore, it minimized human discretion to the point that it negated collective buy-in and created ethical dilemmas where caseworkers felt pressured to manipulate the system. Additionally, each dimension of the framework had cross-dependencies with others, demonstrating practical trade-offs at both the macro- and micro-levels of the model. For example, predictive versus prescriptive types of decision-support affected administration and training in different ways. The predictive nature of CANS tended to replace the need for training for interpreting outcomes (i.e., providing the answer), while the prescriptive nature of 7ei augmented training by helping new CWS employees learn through the collaborative brainstorming process. Thus, a core contribution of this paper is that we demonstrate how the theoretically derived ADMAPS framework (empirically validated through this case study in child-welfare) can be used within other public sector domains (e.g., criminal justice \cite{haque2019exploring, haque2020understanding}) to assess the strengths and weaknesses of algorithms making high-stakes decisions in the lives of people. For instance, ADMAPS can help develop algorithms for judicial decision-making designed to aid judges. Judges must make decisions based on legal justification, interpretation, and application based on relevant laws and precedents \cite{ellsworth2005legal}. The fluid nature of legal reasoning is often at odds with the discrete predictive nature of algorithms \cite{greene2020hidden}. Here, similar to the 7ei algorithm, ADMAPS can help develop analogues that augment human discretion instead of curtailing it. Similarly, ADMAPS can help design algorithms for job placement centers where the caseworkers must exercise discretion in applying complex legal frameworks, assessing resource constraints, as well as resolving organizational contradictions (i.e., \textit{bureaucratic processes}) to extend unemployment benefits to citizens \cite{holten2020shifting}. 


\subsubsection{Holistic Assessments, Not Deterministic Scoring Lead to Improved Decisions}
Through the lens of ADMAPS, the tensions between the three elements (human discretion, bureaucratic processes, and algorithmic decision-making) become apparent. This further implies that policymakers and most practitioners do not get the opportunity to understand and appreciate the value of holistic algorithmic assessments that can augment the current mandates of univariate risk scores. These mandates exist based on state legislation and are meant to provide a legal justification for using algorithms in child-welfare. While the legal implications are out of scope for this paper, our main takeaway is that the current tension that exists between the state and the agency needs to be redressed in order to improve outcomes for children. This is not to say that we must mandate decision-making from both CANS and 7ei but on the other hand, reduce the dependency on singular metrics and increase the dependency on understanding the underlying trauma and behaviors of a particular child. Moreover, as clearly depicted through our results, using singular metrics from a risk assessment to make subsequent decisions about unrelated determinants (for e.g., calculating compensation for foster parents) only incentivises gaming the system and results in a vicious cycle for children in foster care. The benefits of strength-based approaches have been supported in prior literature \cite{badillo2018chibest, pinter2017adolescent, saxena2020human, topitzes2019trauma} as well as the pitfalls for reappropriating algorithmic outcomes for different purposes \cite{saxena2020child, lyons2004measurement}. More importantly, algorithmic decision-making should not be treated as inevitable; knowing when not to design \cite{baumer2011implication}, not to deploy \cite{barocas2020not}, and to resist \cite{saxena2020collective} is equally important. If systems designers and policymakers are unable to balance the dimensions of ADMAPS in a way that it serves on-the-ground practitioners then alternative, non-algorithmic approaches such as collaborative assessments or processes should be developed that support and/or streamline bureaucratic processes and augment the quality of human discretionary work.

\subsection{Design Guidelines for Algorithmic Decision Support Systems in the Public Sector}
Based on our findings we provide the following design guidelines for developing algorithmic decision-support systems in the public sector. Our guidelines highlight the need to support the complex interdependencies between the three dimensions of ADMAPS.

\begin{itemize}[leftmargin=*]
\item Consider making algorithmic outputs multidimensional, rather than a singular metric. This allows for flexibility in interpreting the output through the use of human discretion.

\item Make algorithmic metrics suggestions, rather than mandated decision outcomes. Creating flexibility in the organizational process will reduce bureaucratic overhead and prevent the system from getting overburdened.

\item Account for the degree of uncertainty in the data and the associated outcomes and make room for value judgments and heuristic decision-making.

\item Design algorithmic systems to be used collaboratively so that joint oversight and expertise can provide fairness, transparency, and accountability.

\item Consult key stakeholders to form a consensus on what data should and should not be collected to inform high-stake decisions.

\item Design the system to learn and adapt from expert users by being able to identify exemplar cases and the reasons why they were successful. This may lead to serendipitous data points that were not previously captured formally by the system.

\item Account for the organizational resources and constraints within which all decisions (human and algorithmic) must be made and incorporate this into algorithm design.

\item Avoid direct trade-offs between input variables that create ethical dilemmas. Create safeguards in the system that check for gaming behaviors, such as "what-if" analyses.

\item If tradeoffs cannot be properly managed, consider alternative, non-algorithmic approaches that streamline and/or support bureaucratic processes and augment human discretion.
\end{itemize}


The objective of employing algorithms in the public sector is to improve decision-making by providing efficient, consistent, fair, and defensible decisions. However, as depicted by this case study and the ADMAPS framework, there is a high degree of uncertainty associated with these algorithmic decisions which are further constrained by policies and organizational factors. Therefore, designing for algorithms that support practitioners' decision-making processes (instead of providing predicted outcomes) may offer higher utility and augment the quality of human discretionary work.

\section{Limitations and Future Work}
Our study contributes both a generalizable framework of Algorithmic Decision-Making Adapted for the Public Sector (ADMAPS) and an in-depth ethnographic case study of sociotechnical systems used in the domain of child-welfare. However, there are several limitations that introduce opportunities for scholars to expand upon this research. First, this study solely focuses on the perspective of CWS caseworkers on algorithmic decision-making systems. We believe it is important to also include the perspective of CWS children and families as they are the affected communities of concern. Future research should focus on understanding the perspectives of families and their (lack of) agency with respect to the decisions made about them through the use of algorithmic systems. Second, this study only focused on the algorithms being used in collaborative team settings; however, there might be other decision-tools that caseworkers might use independently in their daily work. For instance, all CWS employees are required to use a comprehensive state mandated data system that has several data-driven visualizations and decision-tools built into it. But as our results indicate, caseworkers lack an adequate understanding of algorithms or decision-tools and only recognize the automated decision-making aspects when explicitly asked to think in those terms. Therefore, future research should focus on studying the state-mandated data systems, in-built decision-tools, and their impact on caseworkers' decisions more holistically.


\section{Acknowledgments}
This research was supported by the National Science Foundation (CRII-1850517) and the William T. Grant Foundation (\#187941, \#190017). Any opinion, findings, and conclusions or recommendations expressed in this material are those of the authors and do not necessarily reflect the views of our sponsors or community partners. We would like to thank our collaborators at the child-welfare agency, SaintA, for allowing us to conduct this extensive ethnography. We are also thankful for our study participants as well as the anonymous reviewers whose suggestions and comments helped improve this manuscript.

\section{Conclusion}
We conducted an in-depth ethnographic study to understand the daily algorithmic practices of caseworkers at a child-welfare agency. Concurrently, we also developed a cohesive framework of Algorithmic Decision-Making Adapted for the Public Sector (ADMAPS) by systematically reviewing and synthesizing prior literature in HCI, STS, and PA. We qualitatively coded our data from the ethnography to the dimensions of ADMAPS to reveal the complex interdependencies between human discretion, algorithmic decision-making, and bureaucratic processes. Our findings show that there is a need to invest in strength-based approaches centered in ecological frameworks. This approach not only seeks to improve the lives of people but also builds collective trust in the outcome itself, and subsequently, leads to collective buy-in at both the caseworker and bureaucratic levels. Moreover, algorithms need to be designed such that they augment human discretion by allowing practitioners to engage in value judgments and heuristic decision-making. In addition, algorithms need to be fully supported by bureaucratic processes by allocating necessary resources and accounting for organizational constraints. As a result of this study, we also propose heuristic guidelines for the design of high-stakes algorithmic decision-making tools in the public sector.

\vspace{0.2cm}

\bibliographystyle{ACM-Reference-Format}
\bibliography{bibliography, LitReviewBIB}

\newpage
\appendix

\section{CODEBOOK FOR RQ1}
\begin{table}[H]
\Small
\begin{tabular}{|>{\raggedright}p{5.5cm}|>{\raggedright\arraybackslash}p{8cm}|}
\hline
\rule{0pt}{1.2em}\cellcolor{mygray3}{\textbf{Algorithms and Associated Outcomes}}              & \rule{0pt}{1.2em}\cellcolor{mygray3}{\textbf{Illustrative Quotation}} \\
\hline
\multicolumn{2}{|l|}{\rule{0pt}{1.3em}{\textbf{Child and Adolescent Needs and Strengths (CANS)}}}  \\
\hline
Compensation Calculations (85\%, n=17)          & \textit{"CANS has become all about the rate with little attention paid to actually helping the child. A foster parent said to me.. this is more like a \$1600 kid"} \\
\arrayrulecolor{mygray2}\hline
Mental Heath Assessment (65\%, n=13)            & \textit{"CANS scores translate to actionable steps in form of mental-health services for the child"} \\
\arrayrulecolor{mygray2}\hline
Level of Foster Care (50\%, n=10)               & \textit{"It recommends the level of foster care for the child, but that decision comes down to the availability of placements"}\\
\arrayrulecolor{black}\hline
\multicolumn{2}{|l|}{\rule{0pt}{1.2em}{\textbf{Seven Essential Ingredients (7ei)}}} \\
\hline
Trauma-informed care (90\%, n=18)               & \multirow{3}{8cm}{\textit{"7ei allows us to center every case in the trauma-responsive model and the scores show that it actually is leading to better placement stability and permanency outcomes"}} \\
\arrayrulecolor{mygray2}\cline{1-1}
Placement Stability and Permanency (85\%, n=17) & \\
 &\\
\arrayrulecolor{black}\hline
\multicolumn{2}{|l|}{\rule{0pt}{1.2em}{\textbf{Anti Sex-Trafficking Response Tool}}} \\
\hline
Monitor case for sex-trafficking (65\%, n=13)   & \multirow{2}{8cm}{\textit{"It helps caseworkers be actively aware of and look for sex-trafficking indicators that we wouldn’t have thought of 10 years ago"}} \\
\arrayrulecolor{mygray2}\cline{1-1}
Report case to HART (65\%, n=13)                &  \\
\arrayrulecolor{black}\hline
\multicolumn{2}{|l|}{\rule{0pt}{1.2em}{\textbf{Legal Permanency Status Tool}}} \\
\hline
Quality of current placement (60\%, n=12)       & \multirow{2}{8cm}{\textit{"It helps us be actively aware of the current status of the placement and systemic barriers, and what we need to do to achieve permanency"}}\\
\arrayrulecolor{mygray2}\cline{1-1}
Identify systemic barriers (50\%, n=10)         & \\
\arrayrulecolor{black}\hline
\end{tabular}
\vspace{0.2cm}
\caption{RQ1: Associated Algorithms and Outcomes. Percentages (\%) represent the proportion of interview participants who recognized and discussed the algorithmic outcome.}
\label{RQ1:codebook}
\end{table}


\section{Participant Profiles}
We interviewed participants who were consistently present at team meetings or who had more experience working with algorithmic systems. While on-site, we interacted with approximately 120 employees and external consultants of the agency. We did not collect demographic information from all meeting participants as we did not want to disrupt their work as it was being carried out.  We did, however, collect the gender, job title, education-level, years of experience of our interviewees, and the algorithms they used in daily work practices as shown in Table \ref{tab:participants}:

\begin {itemize}[leftmargin=*]
    \item \textbf{Child-Welfare Supervisors (n=8)}: Supervise a case management teams consisting of six to eight case managers. All supervisors were former case managers who now provide individual supervision to case managers and ensure their professional development. They facilitate the planning meetings and supervise about 140 cases each.

    \item \textbf{Permanency Consultants (n=5)}: Responsible for permanency planning, which includes the drafting and completion of documents for court and initiating permanency legal process, targeted recruitment for adoption, provision of post-guardianship services and/or post-adoption services for families. On average, they provide direct consultations on about 150 cases each.

    \item \textbf{Child Welfare Case Managers (n=2)}: The front-line workers who directly interact with families and are the point of contact between parents, CWS, and the court system. They conduct home visits, support and monitor the safety and well-being of children, document safety assessments, facilitate improving parental protective capacities by creating goals with families etc. On average, they manage about 20 cases.
    
    \item \textbf{Data Specialists (n=2)}: Track case-level data for families and are also responsible for performance benchmarks at the agency as mandated by federal legislation. Data specialists also analyze data about the algorithms being used at the agency and present findings to agency leadership. 

    \item \textbf{Child Welfare Program Directors (n=1)}: Supervise and manage the child-welfare supervisors. Program directors also manage the training and professional development programs, research projects, as well as policy development initiatives.

    \item \textbf{Permanency Consultation Supervisor (n=1)}: Supervise and manage the permanency consultants and facilitate permanency consultations.

    \item \textbf{Clinical Therapist (n=1)}: Licensed clinical social workers who conduct mental-health assessments for foster children.
\end {itemize}

\vspace{0.2cm}
Since CWS experiences a high turnover rate in the case manager position, we focused our attention more so towards supervisors and permanency consultants when conducting our interviews. Moreover, supervisors facilitate planning meetings and permanency consultants facilitate permanency consultations, which ensured that these staff members were consistently present at all the meetings where collaborative work mediated by policies, social work practice, and algorithms occurred. 
\vspace{0.2cm}

\begin{table}[H]
    \Small
    \begin{tabular}{|>{\centering}p{1.3cm}|>{\centering}p{0.4cm}|>{\raggedright}p{4.5cm}|>{\raggedright}p{2cm}|>{\centering}p{1.3cm}|>{\centering\arraybackslash}p{1.8cm}|}
         \hline
         
         \cellcolor{mygray3}{\textbf{Participant}} & \cellcolor{mygray3}{\textbf{Sex}} & \cellcolor{mygray3}{\centering{\textbf{Job Title}}} & \cellcolor{mygray3}{\textbf{Education}} & \cellcolor{mygray3}{\textbf{Experience (years)}} & \cellcolor{mygray3}{\textbf{Algorithms}}  \\
         \hline
         P1 &  F &  Permanency Consultation Supervisor & MSW, APSW & 22 & 7ei, AST, LPS     \\ 
         P2 &  F &  Permanency Consultant & MSW                    & 20 & 7ei, AST, LPS     \\ 
         P3 &  F &  Permanency Consultant & MSW, APSW              & 9  & 7ei, AST, LPS     \\ 
         P4 &  F &  Permanency Consultant & MSW, APSW              & 8  & 7ei, AST, LPS     \\ 
         P5 &  F &  Permanency Consultant & MSW                    & 12 & 7ei, AST, LPS     \\ 
         P6 &  F &  Permanency Consultant & BSW                    & 3  & 7ei, AST, LPS     \\ 
         P7 &  F &  Child Welfare Program Director & MSW           & 20 & 7ei, CANS         \\ 
         P8 &  F &  Child Welfare Supervisor & MSW, CAPSW          & 19 & CANS, 7ei, AST    \\ 
         P9 &  F &  Child Welfare Supervisor & MSW                 & 13 & CANS, 7ei, AST    \\ 
         P10 & F &  Child Welfare Supervisor & MSW, LPC, NCC       & 9  & CANS, 7ei, AST    \\ 
         P11 & F &  Child Welfare Supervisor & MSW, APSW           & 9  & CANS, 7ei, AST    \\ 
         P12 & F &  Child Welfare Supervisor & MSW                 & 7  & CANS, 7ei, AST    \\ 
         P13 & M &  Child Welfare Supervisor & MSW                 & 12 & CANS, 7ei, AST   \\ 
         P14 & M &  Child Welfare Supervisor & MSW                 & 30 & CANS, 7ei, AST    \\ 
         P15 & F &  Child Welfare Supervisor & MSW                 & 9  & CANS, 7ei, AST    \\ 
         P16 & F &  Clinical Therapist & LCSW                      & 5  & CANS, 7ei, AST   \\ 
         P17 & F &  Child Welfare Case Manager & BSW               & 8  & CANS, 7ei, AST    \\ 
         P18 & M &  Child Welfare Case Manager & BSW               & 2  & CANS, 7ei, AST    \\ 
         P19 & F &  Data Specialist (Program Director) & MSW       & 17 & CANS, 7ei, AST   \\ 
         P20 & M &  Data Specialist & MSW                          & 17 & CANS, 7ei, AST    \\ 
         \hline
    \end{tabular}
    \vspace{0.1cm}
  \begin{tablenotes}
        \item [1]\textbf{BSW}: Bachelor of Social Work \hspace{1.8cm} \textbf{MSW}: Master of Social Work
        \item[2] \textbf{LCSW}: Licensed Clinical Social Worker \hspace{0.8cm} \textbf{LPC}: Licensed Professional Counselor
        \item[3]\textbf{APSW}: Advanced Practice Social Worker \hspace{0.6cm} \textbf{CAPSW}: Certified Advanced Practice Social Worker
        \item[4] \textbf{NCC}: National Certified Counselor
        \item [5]
    \end{tablenotes}
    \caption{Interview Study Participants}
    \label{tab:participants}
\end{table}






\end{document}